\documentclass[twocolumn,floatfix,prb,aps,showpacs]{revtex4-2}
\usepackage{graphicx,amsmath,amssymb,color,nicefrac,multirow, makecell,float,MnSymbol}
\usepackage[thinc]{esdiff}
\usepackage[unicode=true,colorlinks=true,linkcolor=blue,citecolor=blue,urlcolor=blue]{hyperref}
\usepackage{bm,physics,enumerate,booktabs,xcolor,float,ulem}
\usepackage[titletoc,title]{appendix}

\makeatother
\begin{document}
\title{Full distribution and large deviations of local observables in an exactly solvable current carrying steady state of a strongly driven \textit{XXZ} chain}
\author{Sandipan Manna and G. J. Sreejith}
\affiliation{Indian Institute of Science Education and Research, Pune 411008, India}
\begin{abstract}
Current carrying steady states of interacting spin chains exhibit rich structures generated through an interplay of constraints from the Hamiltonian dynamics and those induced by the current. The \textit{XXZ} spin chain when coupled to maximally polarizing Lindblad terms (with opposite signs on either end) admits an exact solution for the steady state in a matrix product state (MPS) form. We use this exact solution to study the correlations and distributions of local spin observables in the nonequilibrium steady state. We present exact expressions for spin correlators, entropy per site and scaled cumulant generating functions (SCGF) for distributions of local observables in the \textit{XX} limit (Ising anisotropy $\Delta=0$). Further, we use the exact MPS solution in the $\Delta>0$ regime, to calculate numerically exact entropy, correlations, as well as full distributions of spin observables in large systems. 
In systems where $\Delta$ is a cosine of rational multiple of $\pi$, we can numerically exactly estimate the large system limit of the SCGF and the large deviation/rate functions of local-$z$ magnetization. 
For these, we show that the deviations of the SCGF, calculated in finite systems, from the asymptotic large system size limit decay exponentially with system size; however, the decay rate is a discontinuous function of $\Delta$.  
The $x$ magnetization density shows a double peak structure at $\Delta\lesssim 1$, suggesting short-range ferromagnetic ordering in the $x$ direction similar to what was reported for the ground state of the \textit{XXZ} chain. 
\end{abstract}
\maketitle
\section{Introduction}

Expectation values of local observables in the long-time limit of typical unitarily evolving, large  quantum manybody systems are described by a Gibbs ensemble or a generalized Gibbs ensemble with chemical potentials determined by the expectation values of the conserved charges in the initial state. Less is known about description of local properties and correlations in steady states of nonequilibrium systems with nonzero charge current except when a hydrodynamic description is available. Local properties in general out-of-equilibrium states of interacting microscopic models are usually difficult to access due to the computational cost involved in producing accurate microscopic representation of the nonequilibrium steady state (NESS).

Dynamics of quantum many-body systems in contact with external reservoirs has been a major focus of both theoretical and experimental investigations in recent years. In the weak coupling regime, Lindblad type master equation~\cite{Lindblad1976,GVKS} can approximate the effective nonunitary dynamics of the system after integrating out the reservoir degrees of freedom. We study a specific case of an \textit{XXZ} spin chain coupled at the ends to spin baths represented by maximally polarizing Lindblad couplings. Despite strong interactions and strong bath coupling that are well outside the linear response regime, the NESS can be exactly represented in MPS form~\cite{prosen2011exact}. Several closely related models also admit exact NESS representations in various forms~\cite{vznidarivc2010exact,vznidarivc2014exact,buvca2014exactly,popkov2020exact,popkov2020inhomogeneous,prosen2014exact,vznidarivc2011solvable}. We bypass the numerical limitations in the usual studies of interacting nonequilibrium systems by using the exact, numerically convenient MPS representation of the NESS presented in Ref.~\cite{prosen2011exact} to describe its features.

In addition to correlations and expectation values of the magnetization, we also calculate the full distribution of the local spin observables in the NESS. The full distribution can be a useful characterizing tool in experimental cold atom systems due to access to the same through time-of-flight methods and snapshot images with single-atom resolution on such platforms~\cite{cherng2007quantum,klich2009quantum,lovas2017full,arzamasovs2019full}. This has inspired studies of full distribution of local observables and current in diverse settings where demonstrative calculations can be performed on microscopic toy models of quantum many body systems such as in 1-D Bose gas under quantum quench~\cite{rylands2019quantum,perfetto2019quench,bastianello2018sinh,arzamasovs2019full}, ground states of \textit{XXZ} chain~\cite{collura2017full}, ground states of 2D Bose Hubbard model~\cite{wang2023distinguishing}, \textit{XXZ} chain after quench~\cite{collura2020order,calabrese2020full,gopalakrishnan2024distinct}, 2D/3D Heisenberg model after quench~\cite{senese2023out}, free fermion and \textit{XY} model ground states~\cite{ares2021exact, najafi2017full}, ground state of Haldane Shastry model~\cite{stephan2017full}, long range interacting Ising model~\cite{ranabhat2024thermalization}, quenched Ising models with dynamical confinement~\cite{tortora2020relaxation} and random circuits~\cite{cecile2024measurement, agrawal2022entanglement}. The integrated current distribution that characterizes dynamical processes has also been calculated in random quantum circuits~\cite{mcculloch2023full}, in the driven \textit{XX} model~\cite{vznidarivc2014exact} and the \textit{XXZ} chain with dephasing baths~\cite{samajdar2023quantum}. 

Present work studies the distribution of local observables in the current carrying steady state of an interacting spin system where this calculation can be reliably performed. Calculations of local properties using the solution for NESS in the MPS form reduces to numerically convenient transfer matrix evaluations. We present explicit expressions of the spin correlations and entropy density as well as the scaled cumulant generating function (SCGF) for the $z$-magnetization density, $z$-domain wall density, and $z$-antiferromagnetic ordering density in the \textit{XX} model (Ising anisotropy $\Delta=0$). We find that the increasing current causes short range antiferromagnetic ordering reflected in the dependence of local spin densities on spin current in the bulk.
 In the finite $\Delta$ setting, numerically exact results in the limit of infinite system size can be obtained when $\Delta$ has the form $\cos \frac{p}{q}\pi$ for co-prime integers $p,q$. We also analyze the effect of finite system size on the SCGF of the $z$-magnetization density. For finite $p$ and $q$, the asymptotic limit is approached exponentially. We show that the rate of approach to the asymptotic large system size limit is a strongly discontinuous function of $\Delta$. The rate of convergence to aymptotic SCGF vanishes with increasing $p$ and $q$. This, for instance, implies that $(p,q)=(1001,4001)$ has a much smaller gap than $(p,q)=(1,4)$ though they have similar values of $\Delta$.

The paper is organized as follows. We introduce the model in Sec.~\ref{sec:Model} and the exact representation of the NESS in terms of MPS in Sec.~\ref{sec:mpsExact}. We provide a brief overview of the proof of the MPS representation of the NESS following Ref.~\cite{prosen2011exact}. This section establishes the notation that we use in the rest of the paper.
In Sec.~\ref{sec:XXmodelResults}, we present the results for the noninteracting (\textit{XY} limit) limit of the model where analytical calculations are possible. We calculate the full distribution of the local observables, the two-point correlations, and the entropy density in the NESS. In Sec.~\ref{sec:XXZmodelResults}, we present the results on full distribution of local magnetization and spin current for the general \textit{XXZ} model at finite Ising anisotropy $\Delta$. We also discuss the symmetries of the NESS and the implications of these symmetries on its properties. Section~\ref{SCGF} discusses rate function (also referred to as large deviation function) for finite systems as well as asymptotic limit of the SCGF. We conclude in Sec.~\ref{sec:conclusion}. Our primary contribution are the results beyond expectation value of local observables, i.e., results on full distribution and large deviation statistics of local observables in a strongly driven interacting spin chain.

\section{Model\label{sec:Model}}
We study a spin-1/2 \textit{XXZ} chain of length $N$ coupled to spin-baths at both ends. The time evolution $\partial_t \rho=\mathcal{L}\rho(t)$ of the density matrix is determined by the Liouvillian $\mathcal{L}$, 
\begin{align}
    \label{Lindblad_equation_def}
    \mathcal{L}\rho = -\imath[H_{XXZ},\rho] +\epsilon \mathcal{D}_1\rho + \epsilon \mathcal{D}_N\rho.
\end{align}
The evolution in the bulk is governed by the Hamiltonian with local density $h_{i,i+1}$
\begin{align}
    \label{def_H_XXZ}
    H_{XXZ} =\sum_i^{N-1} h_{i,i+1} 
            =J\sum_{i}\sigma^x_i \sigma^x_{i+1}+\sigma^y_i \sigma^y_{i+1}+\Delta \sigma^z_i \sigma^z_{i+1} 
\end{align}
where $\Delta$ is the {Ising} anisotropy, $\sigma_{i}^{\alpha}$ are Pauli matrices and $\epsilon$ is the strength of the coupling between the chain and the bath. We will set $|J|=1$ in all calculations.
The dissipator $\mathcal{D}_k\rho$ has the Lindblad form
\begin{align}
\mathcal{D}_k\rho(t)=2 L_k \rho(t) L^\dag_{k} - \{ L^\dag_{k} L_k,\rho(t)\}
\end{align}
acting on spins at the two ends of the chain $k=1,N$. $L_1=\sigma^+$ and $L_N=\sigma^-$ are the jump operators and $\sigma^\pm=\frac{\sigma^x\pm\imath \sigma^y}{2}$. With these, the dissipative terms at the two ends simplify to 
\begin{eqnarray}
{\mathcal{D}_1}\rho_{}=(\sigma_1^{+} \rho_{} \sigma_1^{-}- P^\downarrow_1\rho_{}) + {\rm H.c.,}\nonumber\\
{\mathcal{D}_N}\rho_{}=(\sigma_N^{-} \rho_{} \sigma_N^{+}- P^\uparrow_N\rho_{}) + {\rm H.c.,}\label{eq:simplerDissipator}
\end{eqnarray}
where $P^\nu_k$ represents the projector onto the state $\nu$ acting on site $k$.

\begin{figure}
    \includegraphics[width=\columnwidth]{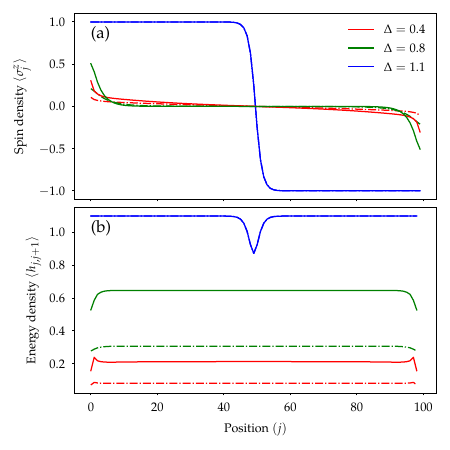}
    \caption{(a) Local $z$-direction magnetization and (b) local energy density as a function of position. The solid lines and dotted lines are for $\epsilon=1$ and $0.5$, respectively. Results for system size $N=100$.\label{fig:spin-energy-density}}
\end{figure}

We are interested in the NESS density matrix $\rho_{\infty}$ which satisfies $\mathcal{L}\rho_{\infty}=0$. The $z$-magnetization density profile and the energy density profile in the NESS are shown in Fig.~\ref{fig:spin-energy-density}. For $|\Delta|<1$, the bulk shows a flat zero magnetization profile in the thermodynamic limit, with a finite magnetization at both ends. Such flat charge density profiles are ubiquitous in the NESS of integrable and ballistic quantum chains~\cite{prosen2009matrix,Nishad2022}.
The energy density profile shows a similar behavior with a jump at the ends and flat profile in the bulk. For $|\Delta|>1$, the spins are close to maximally polarized but along opposite directions on the left and right halves of the chain, with a sharp jump in magnetization at the center. The energy density shows a similar flat profile except near the center. Figure~\ref{fig:spin-energy-density} shows the profiles for $\Delta >0$. 
For $\Delta<0$, the spin profile remains the same but the energy profile is sign-reversed (Also see discussion of symmetries in Sec.~\ref{Symmetry considerations}).

\begin{figure}
    \includegraphics[width=\columnwidth]{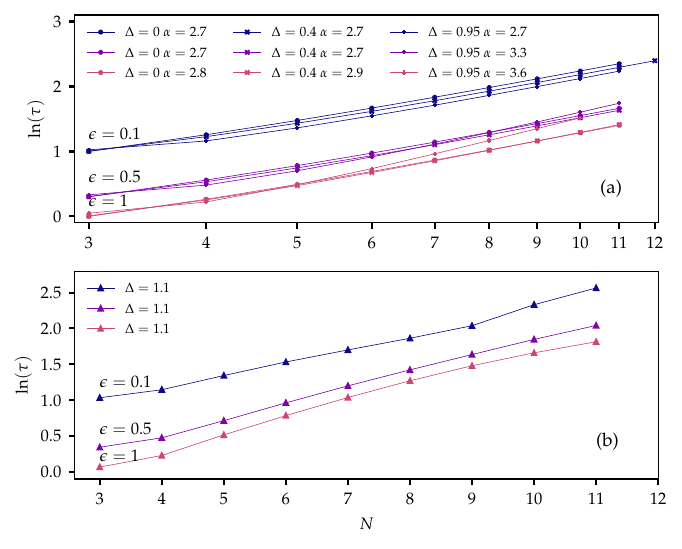}
	\caption{Relaxation time scale $\tau$ as a function of system size. (a) In the gapless regime the relaxation time appears to increase with system size as $\tau \sim N^{3}$. Exponent was obtained by fitting the largest three $N$ values in each case to $\tau\sim N^{\alpha}$. The plot is in log-log scale.(b) In the gapped regime the relaxation time,$\tau \propto \exp(cN)$.  {The plot is in semilog scale.}~\label{fig:timescales}}
\end{figure}
Spin and energy expectation values in the Fig.~\ref{fig:spin-energy-density} are obtained from the exact solutions from Ref.~\cite{prosen2011exact} summarized in Sec.~\ref{sec:mpsExact}. In this work, we use the exact solutions and go beyond the expectation values of local observables and study their correlations and the full distribution in large blocks as well as the entropy in the $\rho_\infty$. 
Calculation of such quantities requires the $\rho_\infty$ for large system sizes with high precision so that longer range and multisite correlations contributing to the full distribution are faithfully captured in the NESS. Direct calculation of $\rho_\infty$ by exact diagonalization (ED) of $\mathcal{L}$ is unfeasible beyond $N \sim 12$ sites. Accurate TEBD estimation is also impractical for systems exceeding $\sim100$ sites, as the time scale required for relaxation to the NESS extends past the regime where tensor network algorithm predictions remain reliable.

Figure~\ref{fig:timescales} shows the rapid growth of relaxation time scale $\tau$ (estimated in small systems using ED of the Liouvillian as the reciprocal of the eigenvalue with the second largest real part) as a function of system size $N$. As was already demonstrated in Ref.~\cite{vznidarivc2015relaxation}, the relaxation time scales with system size as $\tau \propto n^3$ for $|\Delta|<1$ and $\tau \propto \exp(cn)$ for $|\Delta|>1$. The latter scaling is due to the maximally polarized regions in the chain that suppress the dynamics that leads to relaxation. The $\tau$ scales as $\sim n^3$ for all $\Delta$ if the baths are not maximally polarized~\cite{vznidarivc2015relaxation}. While $\Delta$ within each phase has little effect on $\tau$, relaxation time increases with decreasing bath coupling $\epsilon$ through the amplitude of the power-law and exponential scaling.

We take advantage of the access to the exact solution for $\rho_\infty$ of the nontrivial interacting system to calculate the full distribution and rate functions for the fluctuations of the local observables in the bulk of the chain demonstrating the large deviation principle for these quantities. Additionally, we calculate the correlations and entropy density in $\rho_\infty$.

\section{MPS form of the nonequilibrium steady state\label{sec:mpsExact}}
The NESS $\rho_\infty$ can be exactly expressed in a matrix product form as was shown in Ref.~\cite{prosen2011exact,PhysRevLett.106.217206}. We first state the solution. We also provide a short summary of the proof after this for completeness. We will need only the solutions shown in Eq. [\ref{definition_Sn}, \ref{eq:Amatrices}, \ref{eq:AppendixAmatrixElements}] in the rest of the paper. In the subsequent subsection, we also illustrate the construction of MPO for local observables and the corresponding transfer matrix representations.

\subsection{MPS Solution}
The NESS $\rho_{\infty} $, being a positive semidefinite matrix, can be expressed as a product $S_N S_N^\dagger$. An ansatz for $S_N$ can be written as a matrix product state (MPS) over Pauli strings of the following form:
\begin{eqnarray}
    &S_N = \sum \psi^{s_1s_2s_3\dots} \times
      \sigma^{s_1}\otimes \sigma^{s_2}\otimes\dots \sigma^{s_N}, \label{definition_Sn}\\
      &\psi^{s_1s_2s_3\dots}=\langle 0 |\prod_{i=1}^NA_{s_i}|0\rangle. \nonumber
\end{eqnarray}
Here, $s_i$ takes values from $\{+,-,0\}$, $\sigma^{s_i}$ acts on the $i^{\rm th}$ site and $\sigma^0=\mathbb{I}/\sqrt{2}$. The local tensors $A_{s}$ act on the bond Hilbert space $H_B$ and have the following form
\begin{align}
A_0 &= \sqrt{2}|0\rangle \langle 0 | + {{\sum}}_{r=1}^{\infty}  \sqrt{2}a_r^0 |r\rangle \langle r|,  \nonumber \\
A_+ &= \imath \epsilon |0\rangle \langle 1 | + {{\sum}}_{r=1}^{\infty} a_r^+ |r\rangle \langle r+1|,  \label{eq:Amatrices}  \\
A_- &= |1\rangle \langle 0 | + {{\sum}}_{r=1}^{\infty} a_r^- |r+1\rangle \langle r|.\nonumber
\end{align}
$H_B$ is spanned by the orthonormal states $|r\rangle$, $r=0,1,\dots$. The explicit form of the matrix elements are given in the appendix Eq.~\ref{eq:AppendixAmatrixElements}. 

We briefly comment on the specific structure of the ansatz. Firstly, while the most general ansatz for $S_N$ would contain $\sigma_z/\sqrt{2}$ in addition to $\{\sigma^{\pm}, \sigma^0\}$, $S_N$ does not contain $\sigma_z$.
Secondly, symmetries of $\mathcal{L}$ implies that the density matrix commutes with the spin operator $S_z=\sum_{i}\sigma^z_i$, i.e $[S_z,S_NS_N^\dagger]=[S_z,S_N]S_N^\dagger-{\rm H.c.}=0$. The above ansatz assumes specifically that $[S_z,S_N]=0$. This local symmetry of $S_N$ suggests that it can be written in terms of quantum number conserving tensors $A$ (see for instance, Ref.~\cite{Schollwck2011}) where each bond index is associated with a charge. The tri-diagonal structure in the above ansatz assumes that the dimension of each charge sector is $1$ implying the states can be labeled by the charge. The symmetry $[S_z,S_N]=0$ implies that the symmetry charge of the bond state appearing on the left and right ends are the same and can be chosen to be $0$. Last, boundary conditions imply that $A_-|0\rangle=0$ which suggests that negative bond charges are not needed. 

\subsection{Summary of the proof}
\begin{figure}[h!]
    \includegraphics[width=1\columnwidth]{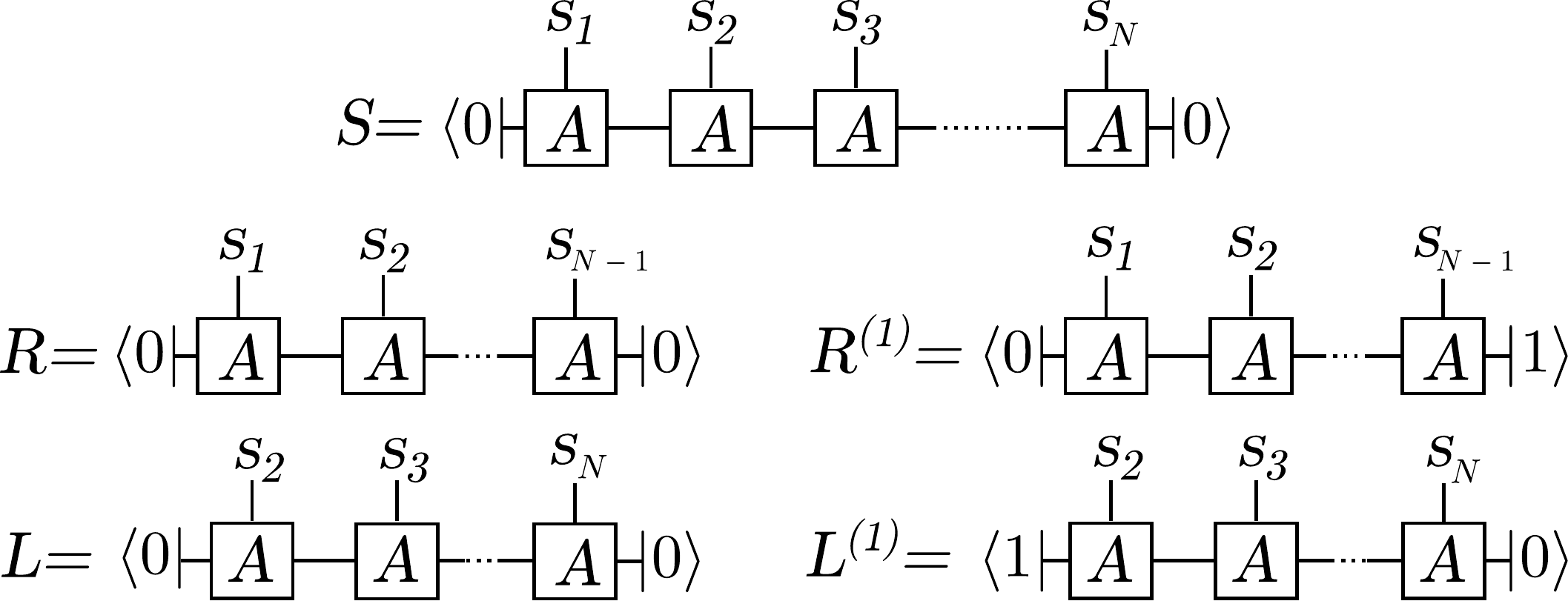}
    \caption{MPS representation of $L,L^{(1)}$ and $R,R^{(1)}$ acting on sites $1<i\leq N$ and $1\leq i<N$ respectively. The operators $L^{(1)}$ and $L$  differ in the bond charge on the left end. Similarly $R^{(1)}$ and $R$ differ in the bond charge at the right end. The leg indices are Pauli indices $s_i$. \label{MPO_diagram}}
\end{figure}
Before sketching the proof, it is useful to define the operators shown in Fig.~\ref{MPO_diagram} in their MPS form. Operators $L$ and $R$ are equivalent to $S_{N-1}$ but act on the $N-1$ sites $2,3\dots N$ and $1,2,\dots N-1$ respectively. Operators $R^{(1)}$ and $L^{(1)}$ are similar to $R$ and $L$ respectively but unlike them these are terminated on the left and right ends respectively by $|1\rangle$ instead of $|0\rangle$.

We first state a sufficient condition on $S_N$ such that $S_NS_N^\dagger$ is the steady state. If $S_N$ can be written as
\begin{align}
    S_N &= \mathbb{I}_1 \otimes L + \imath \epsilon \sigma^+_1 \otimes L^{(1)}\nonumber\\
    S_N &= R\otimes \mathbb{I}_N +   R^{(1)}\otimes \sigma^-_N\label{Eq:SRLrelations}
\end{align}
and it satisfies the following divergence condition 
\begin{align}
\label{eq:divergence_condition}
    [H,S_N]=-\imath \epsilon(\sigma_1^z \otimes L - R\otimes \sigma_N^z)
\end{align}
then $\rho_\infty=S_NS_N^\dagger$ is annihilated by the $\mathcal{L}$.  

To see this, we consider the action of the dissipator, on the density matrix. Using Eq.~\ref{eq:simplerDissipator} in Eq.~\ref{Eq:SRLrelations}, we get 
\begin{align}
&\epsilon {\mathcal{D}_1}S_NS^\dagger_N =\epsilon(\sigma_1^z \otimes LL^{\dag} + \imath \epsilon \sigma^{-}_1 \otimes LL^{{(1)}^{\dag}}) +{\rm H.c..}\nonumber\\
&\epsilon {\mathcal{D}_N}S_NS^\dagger_N =-\epsilon(\sigma_N^z \otimes RR^{\dag}  + \sigma^{+}_N \otimes RR^{{(1)}^{\dag}}) +{\rm H.c..}.\label{dissipatorExpansion}
\end{align}
By identifying terms similar to Eq.~\ref{Eq:SRLrelations} in the ight-hand side (RHS) we see that the first and second lines are $\epsilon \sigma^z_1 L S_N^\dagger + \rm{H.c..} $  and $-\epsilon S_NR^\dagger \sigma_N^z + {\rm H.c..} $, respectively and therefore 
\[
\epsilon (\mathcal{D}_1+\mathcal{D}_N)S_NS^\dagger_N = \epsilon (\sigma^z_1\otimes  L -  R\otimes \sigma_N^z)S_N^\dagger + {\rm H.c..}
\]
This has to be canceled by $-\imath[H,S_NS_N^\dagger]$ so that $\mathcal{L}S_NS^\dagger_N=0$. Using the Leibnitz rule for the commutator,
\[
\imath [H,S_NS_N^\dagger] = \imath [H,S_N]S_N^\dagger + {\rm H.c..}
\]
Comparing the RHS of the above equations we see that Eq.~\ref{eq:divergence_condition} (in addition to Eq.~\ref{Eq:SRLrelations}) is a sufficient condition $S_NS_N^\dagger$ to be the steady state. Solutions to these are presented in Appendix~\ref{sec:Asolutions}.

\subsection{Expectation value as transfer matrix \label{Expectation value from transfer matrix}}
The expectation value of an operator $O$ is given by 
\[\langle O \rangle = {\rm Tr}(S_N^{\dag}O S_N)/Z\]
where $Z={\rm Tr}(S_N^{\dag} S_N),$. Expectation value of a single site operator $O$ on site $k$ can be written as $\langle 0| T^{k-1}V_OT^{n-k}|0\rangle/Z$ where $|0\rangle\equiv|0,0\rangle\equiv|0\rangle\otimes |0\rangle\in H_B\times H_B$, the transfer matrix $V_O$, which maps state $jj'\in H_B\times H_B$ to the state $ii'\in H_B\times H_B$, is given by 
\begin{figure}[h!]
	\includegraphics*[width=0.7\columnwidth]{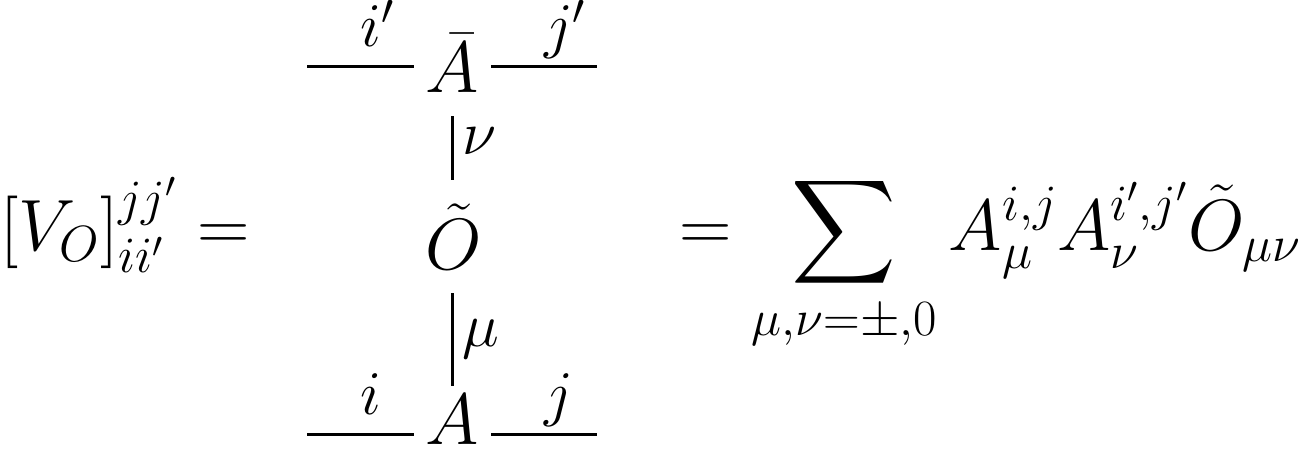}
\end{figure}

\noindent{and} $T=V_{\mathbb{I}}$. The normalization is $Z=\langle 0| T^N|0\rangle$. The $4\times 4$ matrix $\tilde{O}$ is given by
\[
\tilde{O}_{\mu\nu}  = {\rm{Tr}}[\sigma_\mu^\dagger  O \sigma_\nu]\;
\]
where $\sigma_\mu,\sigma_\nu \in \{\mathbb{I}/\sqrt{2},\sigma^-,\sigma^+,\sigma^z/\sqrt{2}\}$. 
Expectation values of direct product of single site operator $O_jO'_{j+1}$ can be calculated as $\langle 0| T^{j-1}V_{O_j}V_{O'_{j+1}}T^{N-j-1}|0\rangle/Z$.

Now we consider the case of more general operator $O$, which cannot be written as a direct product of single site operators. Such an observable has an MPO representation
\begin{figure}[h!!]
	\includegraphics[width=0.7\columnwidth]{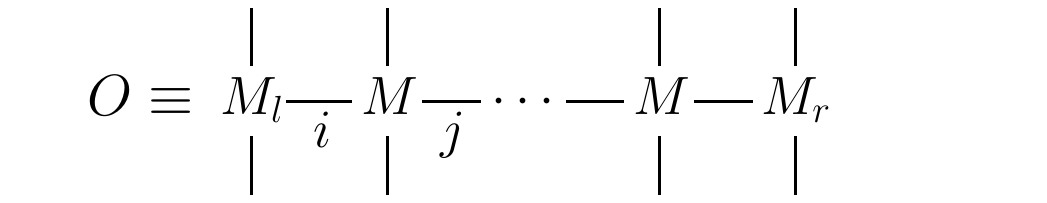}
\end{figure}

\noindent{where the elements $M_{ij}$ of the tensors (for each bond index pair $i,j$) are single-site operators.} The expectation value of $O$ acting on sites $a$ to $a+l$ can be written as 
\begin{align}
	\label{expectation_val_transfermat}
	{\rm Tr}[\rho O] = \langle 0 |  T^{a-1}V_{M_a} V_M^{l-1}V_{M_{a+l}} T^{N-a-l}|0\rangle /Z,
\end{align}
where $V_M$ is obtained by $[V_M]_{ij} = V_{M_{ij}}$ for the single site operator $M_{ij}$, $V_{M_{ij}}$ was defined earlier in this section. 

We now illustrate this with a specific example of the operator corresponding to the imaginary field generating function of the $z$-domain wall count in a block from $a$ to $a+l$. This can be written as $\mathcal{G}(\lambda)={\rm {Tr}}[\rho O]$ where $O=\exp(\imath \frac{\lambda}{2} \sum_{i=a}^{a+l-1}\sigma_i^z\sigma_{i+1}^z)$ up to a phase factor which we ignore for this discussion. $O$ can be expressed as an MPO with with local tensors of bond dimension $2$ as $O=M_a M^{l-1} M_{a+l}$ where
\begin{align}
	\label{MPO Domain wall}
	M_{a}=&\left[
	\begin{array}{cc}
		\mathbb{I} & \sigma^z \\
	\end{array}
	\right] \quad
	M_{a+l}=\left[
	\begin{array}{c}
		\mathbb{I} \cos\frac{\lambda}{2}  \\
		\imath \sigma^z \sin\frac{\lambda}{2}  \\
	\end{array}
	\right] ,\nonumber\\ 
	M&=\left[
	\begin{array}{cc}
		\mathbb{I} \cos\frac{\lambda}{2}  & \sigma^z \cos\frac{\lambda}{2}  \\
		\imath \sigma^z \sin\frac{\lambda}{2}  & \imath  \mathbb{I} \sin\frac{\lambda}{2} \\
	\end{array}
	\right].  \quad  	
\end{align}
Operators here are multiplied as tensor product. We obtain the corresponding transfer matrices as the following block matrices
\begin{align}
	V_{M_a}=&\left[
	\begin{array}{cc}
		T & V_{\sigma^z} \\
	\end{array}
	\right] \quad
	V_{M_{a+l}}=\left[
	\begin{array}{c}
		T \cos\frac{\lambda}{2}  \\
		\imath V_{\sigma^z}  \sin\frac{\lambda}{2}  \\
	\end{array}
	\right] \nonumber\\ 
	V_M&=\left[
	\begin{array}{cc}
		T \cos\frac{\lambda}{2} &  V_{\sigma^z} \cos\frac{\lambda}{2} \\
		\imath V_{\sigma^z} \sin\frac{\lambda}{2}  & \imath T \sin\frac{\lambda}{2} \\
	\end{array}
	\right].  \quad \nonumber
\end{align}
With these, the generating function is given by 
\begin{equation}
\mathcal{G}(\lambda) = \langle 0 | T^{a-1}V_{M_a} V_M^{l-1} V_{M_{a+l}} T^{N-a-l}|0\rangle/Z.
\end{equation}

The state $S_N$ has a maximum bond dimension of $N/2$. For a generic operator $O$ of MPO bond dimension $\chi$, transfer matrices $T,V_{M_l}, V_{M}$ and $V_{M_r}$ in general have sizes $N^2/4\times N^2/4$, $N^2/4\times N^2\chi/4$, $N^2\chi/4\times N^2\chi/4$ and $N^2\chi/4 \times N^2/4$ respectively. For large systems multiplication of these matrices can be prohibitively expensive without suitable truncation of their dimensions.

However, in the case of operators $O$ that can be written as a polynomial in $\{\sigma_z,\mathbb{I}\}$ such as $\mathcal{G}$ introduced before, the evaluation is made efficient by the fact that $\tilde{O}_{\mu\nu}\propto \delta_{\mu\nu}$, and the transfer matrices $V_{\sigma_z}$ and $T$ conserve the quantity $Q(i,i'\in H_B\times H_B)=i-i'$ in the bond indices:
\begin{equation}
	[V_{\sigma_z}]_{ii'}^{jj'} = T_{ii'}^{jj'} = 0 \text{ if }Q(i,i')\neq Q(j,j')
\end{equation}
Since $Q(0,0)=0$ at the two ends and $Q$ is conserved by $T$ and $V$, only the $Q=0$ sector is relevant in every bond and $T,V$ can be restricted to the $Q=0$ sector. This restriction reduces the dimension of the transfer matrices in the relevant sector ($Q=0$) to ${N}/2$ for $T$ and $N\chi/2$ for $V_M$. Calculations using such small transfer matrices can be efficiently performed to high precision due to the $O(N)$ scaling of the matrix sizes.
  
For the case of $O=\sigma_z$ and $O=\mathbb{I}$, $\tilde{O}\equiv {\rm diag}(0,-1,1,0)$ and ${\rm diag}(1,1,1,1)$ respectively. The transfer matrices $V_{\sigma_z}$ and $T=V_{\mathbb{I}}$, restricted to the $Q=0$ sector are given by
\begin{align}
&V_{\sigma_z}= \sum_{r=0}^{\infty} {|a_r^+|^2 }  |r \rangle \langle r+1| -  {|a_r^-|^2}  |r+1 \rangle \langle r| \label{eq:vsigmaz}\\
&T= \sum_{r=0}^{\infty} {|a_r^+|^2 }  |r \rangle \langle r+1| +  {|a_r^-|^2}  |r+1 \rangle \langle r| +  {2|a_r^0|^2 }  |r \rangle \langle r|\label{eq:t}
\end{align}
where $|r\rangle$ here denotes the $|r,r\rangle\in H_B\times H_B$.

\section{Results for \texorpdfstring{$\Delta=0$}{\textit{XX} limit}\label{sec:XXmodelResults}}
In the \textit{XX} limit ($\Delta=0$), the tensors $A$ can be truncated to $2\times 2$ matrices,
\begin{equation}
A^0 =  \sqrt{2}\begin{bmatrix}
		1 & 0 \\[0.3em]
		0 & \frac{\imath \epsilon}{2}
	\end{bmatrix},\;	
A^+ =  \begin{bmatrix}
		0 & \imath \epsilon \\[0.3em]
		0 & 0
	\end{bmatrix},\;
A^-=  \begin{bmatrix}
		0 & 0 \\[0.3em]
		1 & 0
	\end{bmatrix}.\nonumber
\end{equation}
The truncation is due to the fact that the matrix elements in $A$ that take the state out of the space spanned by $\{|0\rangle,|1\rangle\}$ are 0 (See Appendix \ref{Matrix elements of A}).
The transfer matrix in the bond charge sector $Q=0$ (Eq.~\ref{eq:t}) becomes,
\begin{equation}
	T = \begin{bmatrix}
		2 & {\epsilon^2} \\[0.3em]
		1 & \frac{\epsilon^2}{2}
	\end{bmatrix}\nonumber
\end{equation}
which acts on the space spanned by $|0,0\rangle, |1,1\rangle$ in the $Q=0$ subspace of $H_B\otimes H_B$.
The truncation of the dimensions of the $T$ and $V_{\sigma_z}$ allows us to obtain exact closed form expressions for various quantities. In this section, we calculate the properties of local spin operators in the \textit{XX} limit.
\paragraph*{Mean and two-point correlations of spins:}
The transfer matrix $V_{\sigma_z}$ in the $Q=0$ sector (Eq.~\ref{eq:vsigmaz}) is given by, 
\begin{equation}
	V_{{\sigma_z}}=\left[
		\begin{array}{cc}
			0 & {\epsilon ^2} \\
			-1 & 0 \\
		\end{array}
		\right].	\nonumber
\end{equation}
The magnetization $\langle \sigma_z \rangle $ on the $i$th site can be calculated as $\langle T^{i-1}V_{\sigma_z}T^{N-i}\rangle/Z$.
This is exactly $0$ except at the two ends where the magnetization increases monotonically with $\epsilon$ as  $\pm \epsilon^2/(\epsilon^2+4)$.

The two-point $\sigma_z$ correlator can be calculated as
\begin{equation}
	\langle \sigma_z^i \sigma_z^j \rangle = {\langle 0 |  T^{i-1}V_{\sigma_z}T^{j-i-1} V_{\sigma_z} T^{N-j}|0\rangle}/{Z}\nonumber
\end{equation}
The connected two point correlator is zero except on adjacent sites when it is $- 4\epsilon^2/(\epsilon^2+4)^2$\cite{vznidarivc2011solvable}.

Now we consider the magnetization in the $x$ direction. The density matrix commutes with $S_z$ and therefore magnetization $\langle\sigma^{x}\rangle$ is 0 everywhere. The two point correlator $\langle \sigma^x_i \sigma^x_j \rangle$ can be calculated as $2{\rm Re}\langle \sigma^+_i \sigma^-_j\rangle$. We can calculate $\langle \sigma^+_i \sigma^-_j \rangle$ using
\begin{align}
	\langle \sigma^+_i \sigma^-_j \rangle = {\langle 0 |  T^{i-1}V_{\sigma^+}T_{+-}^{j-i-1} V_{\sigma^-} T^{N-j}|0\rangle}/{Z}.\nonumber
\end{align}
Note that $V_{\sigma^\pm}$ alters the $Q$ by $\pm 1$. $T_{+-}$ is the transfer matrix in the $Q=1$ subspace of $H_B\times H_B$. The Q = 1 subspace implies that the bonds can be in a one-dimensional space spanned by the state $|0,1\rangle$. This is because the matrix elements of $A$ that take the bond state out of this space vanish. These are given by 
\begin{align}
T_{+-}= -\imath {\epsilon}/2,\quad 
V_{\sigma^+} = 
	-\imath\epsilon\left[\begin{array}{c}  
		  1 \\
		 { 1}/{2} \\
	\end{array}\right],\quad
V_{\sigma^-} =
\left[
\begin{array}{cc}
	1 & \frac{\epsilon ^2}{2}
\end{array}
\right].\nonumber
\end{align}

The two point correlation in the bulk can be calculated to be 
\begin{align}
	\label{sigx_sigx_exact_form}
\langle \sigma^x_p \sigma^x_{p\pm k} \rangle =4\left(\frac{2\epsilon}{\epsilon^{2}+4}\right)^{k}\cos\frac{\pi k}{2},
\end{align}
assuming $p,p\pm k$ are not the ends $1$ or $N$. 

The correlations are $0$ for all odd values of the distance $k$ and otherwise is oscillatory with an exponentially decaying amplitude $\sim [{2\epsilon}/(4+\epsilon^2)]^k$. Local spin current expectation value can be calculated as $2 J{\rm Im} \langle \sigma^+_i \sigma^-_{i+1}\rangle$ and gives
\begin{equation}
\langle J^s \rangle = \frac{2\epsilon}{4+\epsilon^2}J.\label{eq:currentExpect}
\end{equation}

\begin{figure}
	\includegraphics[width=\columnwidth]{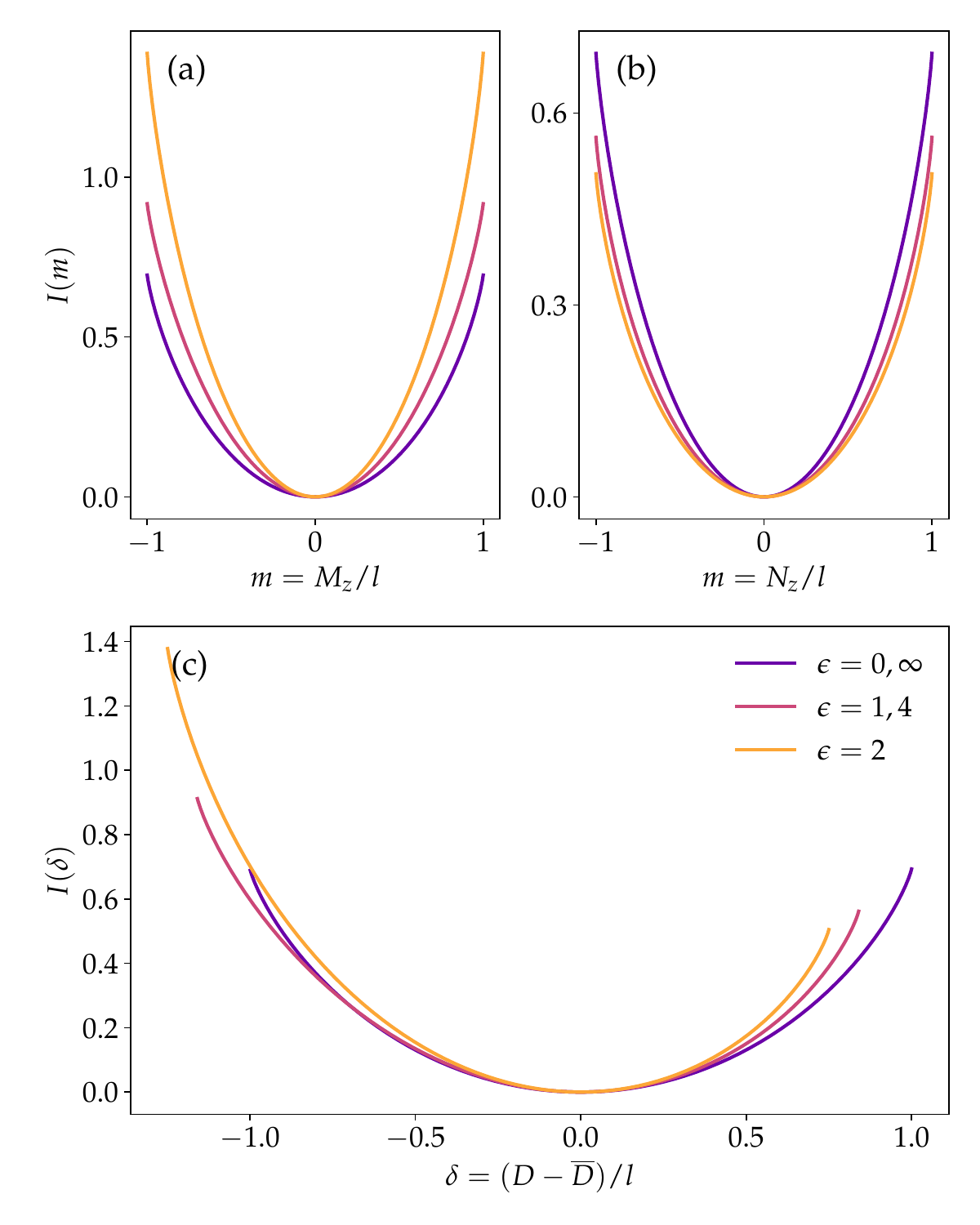}
	\caption{Rate functions for the (a) magnetization, (b) staggered magnetization and (c) domain wall density in the NESS of the \textit{XX} model obtained from Legendre transform of Eq.~\ref{SCGFBlock}, Eq.~\ref{SCGFStagg} and Eq.~\ref{SCGFDomainWall} for different representative values of $\epsilon$. Rate functions are the same at $\epsilon$ and $4/\epsilon$.
	\label{fig:Delta0Ratefunction}}
\end{figure}

\subsection{Distribution of local observables in a block}
\paragraph{$z$-magnetization}
We now study the distribution of the local observables in the system. The total magnetization in a block of size $l$ between $a$ and $a+l$ is given by $(N=2n+l)$,
\begin{equation}
M^{z}_l=\sum_{j=a+1}^{a+l} \sigma^{z}_j\nonumber
\end{equation}
$M^{z}_l$ has eigenvalues $M \in [-l,l]$. The imaginary field generating function $\mathcal{G}_{M_l^{z}}=\langle e^{\imath \lambda M_l^{z}} \rangle$ for the magnetization distribution can be written as $\langle 0 | T^n V_{ez}^l T^n  |0 \rangle/Z$ where (in the $Q=0$ sector)
\begin{align}
	V_{ez} (\lambda)=T\cos \lambda +  \imath V_{\sigma_z}\sin \lambda = \begin{bmatrix}
		2\cos \lambda  &  e^{i \lambda } \epsilon ^2 \\
		 e^{-i \lambda }& \frac{1}{2} \epsilon ^2 \cos \lambda  \\
	\end{bmatrix} \nonumber
\end{align}
The generating function $\mathcal{G}_{M_l^{z}}$ in the {\it bulk} is independent of number of sites on the left and right of the block and can be calculated to be 
\begin{align}
	\mathcal{G}_{M_l^{z}} (\lambda)&= \frac{\alpha_+^{l+1}-\alpha_-^{l+1}}{\alpha_+-\alpha_-} \\
	\alpha_{\pm}&=\frac{1}{2}\left(\cos\lambda \pm \sqrt{\cos^2\lambda + \left(\frac{4\epsilon}{4+\epsilon^2}\right)^2 \sin^2 \lambda}\right)\nonumber
\end{align}
The distribution of magnetization can be obtained as the Fourier transform of $\mathcal{G}_{M_l^{z}}$. 
$M^{z}_l$ at $\epsilon=0$ and $\epsilon\to \infty$ has the binomial distribution $P(M)=C^{l}_{(l+M)/2}$ (where $C^a_b=\frac{a!}{b!(a-b)!}$ are binomial coefficients) as expected for uncorrelated spins in the infinite-temperature ensemble. At $\epsilon=2$, the distribution is given as $P(M)=C^{2l+2}_{l+1-M}$ where $\{M\}$ is the set of eigenvalues of $M_l^{z}$. For general $\epsilon$, the distribution has zero mean and the variance is  
\begin{equation}
	{\rm var}(M^z_l) = \left(l+1\right) \frac{8\epsilon^2}{(4+\epsilon^2)^2}- l.\label{eq:szvariance}
\end{equation}

The generating function along the real counting field $\lambda$ can be calculated in a similar way. This lets us obtain the SCGF as follows,
\begin{multline}
	\label{SCGFBlock}
	\Phi_{M^{z}_l} (\lambda) = \lim_{l\to \infty }\frac{\ln \langle e^{\lambda M^{z}_l}\rangle}{l} \\
	= \ln\frac{1}{2}\left(\cosh \lambda + \sqrt{\cosh^2 \lambda - \left(\frac{4\epsilon}{4+\epsilon^2}\right)^2 \sinh^2 \lambda}\right)
\end{multline}
suggesting that the distribution satisfies the large deviation principle i.e. $P(m) \approx e^{-lI(m)}$ where $I(m)$ is the large deviation rate function and $m=M_l^{z}/l$ is the magnetization density. The rate function $I_\epsilon(m)$ of the magnetization density can be obtained through a Legendre transformation of $\Phi$ (assuming G\"artner-Ellis theorem\cite{gartner,ellis1984large}). The rate function can be explicitly computed at $\epsilon=0,\infty,2$:
\begin{align}
&I_{\epsilon=0,\infty}(m)=H({(1+m)}/{2})\nonumber\\
&I_{\epsilon=2}(m)=2H({(1+m)}/{2})-2H(1/2)\nonumber
\end{align}
where $H$ is the binary entropy function. For general $\epsilon$, the rate function can be obtained by numerical Legendre transformation. The rate function (shown for representative values of $\epsilon$ in Fig.~\ref{fig:Delta0Ratefunction}(a)) interpolates between the above two functions for other $\epsilon$. The distribution is qualitatively the same at all $\epsilon$, it is narrowest at $\epsilon=2$ and broadens on either side of $2$. 

\paragraph{Staggered magnetization}
Generating function for the staggered magnetization distribution in a block of $2l$ spins $N^{z}_{2l} =\sum_{j=x+1}^{x+2l}(-1)^j \sigma^{z}_j$ can be computed as $\langle T^n V_{en}^{l} T^n \rangle/Z$ where $V_{en}(\lambda)=V_{ez}(\lambda) V_{ez}(-\lambda)$. 
This can be explicitly computed to see that the mean is $0$ and the variance of the distribution is given by
\begin{align}
{\rm var}(N^z_l) = \left(\frac{4\epsilon}{4+\epsilon^2}\right)^2 \left(l-1/2\right) + 2l
\end{align}
The SCGF for staggered magnetization can be calculated as the following.
\begin{multline}
	\label{SCGFStagg}
	\Phi_{N^{z}_l}(\lambda) = \frac{1}{2}\log \frac{1}{2}\left(\cosh^2(\lambda) + \frac{8 \epsilon^2 \sinh^2(\lambda)}{\left(\epsilon^2 + 4\right)^2} \right.\\
	\quad \left. + \cosh(\lambda)\sqrt{\cosh^2(\lambda) + \frac{16 \epsilon^2 \sinh^2(\lambda)}{\left(\epsilon^2 + 4\right)^2}} \right)
\end{multline}
The rate function for the staggered magnetization density can be found from numerical Legendre transformation. This is shown in Fig.~\ref{fig:Delta0Ratefunction}(b). The distribution is broadest at $\epsilon=2$ but narrows for values of $\epsilon$ away from this.

\begin{figure}
	\includegraphics[width=\columnwidth]{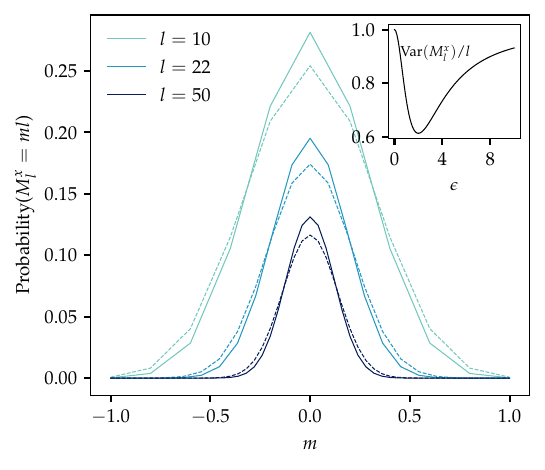}
	\caption{Distribution of $M^x_l$ at $\Delta=0$ for $\epsilon=1$ (solid line) and $10$ (dashed line) The three different colors show different block sizes $l$. (Inset) Dependence of Var($M^x_l$) on $\epsilon$. $N=100$.
		\label{Sx_distribution_D_0_eps_1_10.pdf}}
\end{figure}
	
\paragraph{Domain wall count:} 
Here, we calculate the distribution of domain wall count as \[D^{z}_l=\frac{1}{2}\sum_{j=x+1}^{x+l-1}1-\sigma^z_{j} \sigma^z_{j+1}.\] in a block of size $l$. The generating function can be obtained using the MPO form described in Eq.	\ref{MPO Domain wall}.
\begin{align}
\label{ generating funtion domain wall}
\langle e^{\lambda D^{z}_l } \rangle&= e^{\frac{\lambda}{2} l} \frac{\zeta_++\zeta_-}{\zeta_+-\zeta_-}\left( \frac{\zeta_+^{l+1}-\zeta_-^{l+1}}{\zeta_++\zeta_-}+\zeta_+^{l}-\zeta_-^{l} \right) 
\end{align}
where $l$ is the number of bonds in the block and 
\begin{align}
\zeta_\pm = \frac{1}{2}\cosh \frac{\lambda}{2} \pm \frac{1}{2}\sqrt{\cosh^2 \frac{\lambda}{2} + \left(\frac{4\epsilon}{4+\epsilon^2}\right)^2 \cosh \frac{\lambda}{2} \sinh \frac{\lambda}{2}}  \nonumber
\end{align}
The distribution has a mean and variance
\begin{eqnarray} 
	&\overline{D^{z}_l} = l(1-\frac{4\epsilon }{4+\epsilon^2})\nonumber\\
	&{\rm var}(D^{z}_l)=\frac{1}{8}\left(\frac{4\epsilon}{4+\epsilon^2}\right)^4\left(1-\frac{3l}{2}\right)  + l
\end{eqnarray}
At large-$l$ limit, the SCGF for domain wall count is given by,   
\begin{equation}
\Phi_{D^{z}_l} =\lim\limits_{l \to \infty}\frac{\ln \langle e^{\lambda D^{z}_l } \rangle}{l}=\ln \zeta_+ \label{SCGFDomainWall}
\end{equation}
The rate function for the deviation $D^{z}_l-\overline{D^{z}_l}$ is shown in Fig.~\ref{fig:Delta0Ratefunction}(c). 

\begin{figure}
\includegraphics[width=0.9\columnwidth]{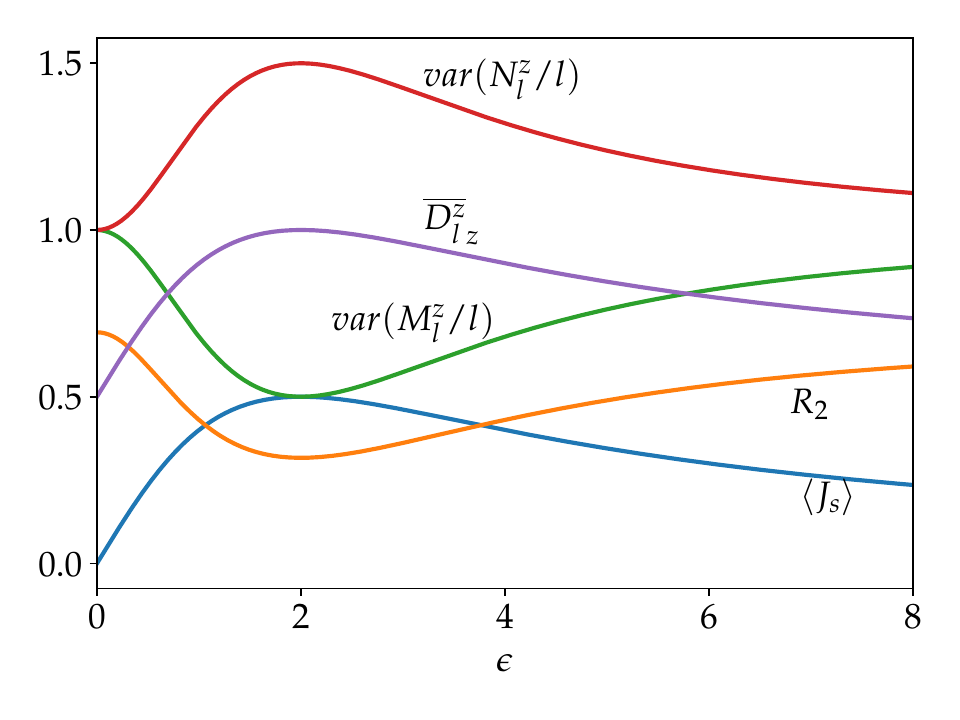}
\caption{Current as a function of the coupling $\epsilon$ shown along with bipartite entropy, variance of the distribution of block magnetization, staggered magnetization and mean domain wall count in the \textit{XX} model.
\label{fig:delta0summary}}
\end{figure}

\paragraph{$x$-Magnetization:} The generating function for $x$-magnetization can also be calculated using transfer matrices but of a larger dimension $4 \times 4$. We could not analytically construct powers of these larger transfer matrices and therefore the SCGF could not be obtained in a closed form. However these can be numerically multiplied to obtain the distribution. Figure~\ref{Sx_distribution_D_0_eps_1_10.pdf} shows the numerically obtained distribution at different values of $\epsilon$. The variance of the distribution is qualitatively similar to that of the $z$-magnetization. Figure \ref{fig:delta0summary} summarizes the variance of different local observables discussed in this section.

\subsection{Bipartite entropy at \texorpdfstring{$\Delta=0$}{Delta=0}}
Last, we calculate the second Renyi entropy in a system bi-partitioned into $N_A$ and $N_B=N-N_A$ sites, $R_2=-\frac{1}{N_A}\ln [({\rm Tr}_B[\rho])^2]$ which can also be computed with a generalization of the transfer matrix method. The entropy per site is given by, 
\begin{equation}
R_2=-\frac{1}{N_A}\ln \frac{\langle 0 | \mathbb{X}^{N_A} \mathbb{T}^{N_B} | 0\rangle }{ \langle 0 | \mathbb{T}^{N} | 0\rangle}.
\end{equation}
Here, $|0\rangle  \equiv | 0,0 \rangle $. We represent the tensor diagrams necessary to evaluate the expression for $R_2$ below in Fig.\ref{entropy_diagram}. The figure illustrates calculation of $R_2$ for the reduced density matrix of first two sites in a system consisting of four sites.
\begin{figure}[H]
\includegraphics[width=0.9\columnwidth]{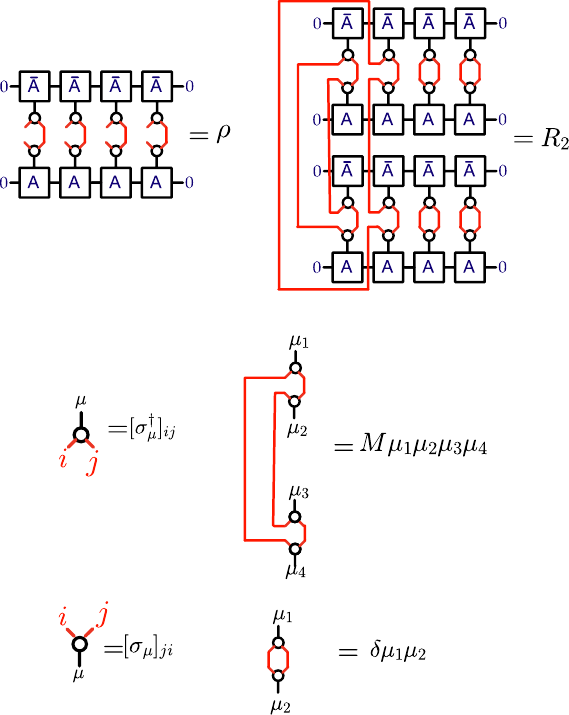}
\caption{ The tensor diagrams that are evaluated to obtain Bipartite Renyi entropy. Density matrix $\rho$ in terms of $S_N$, expression of $R_2$ and $M$ (see the expression for $\mathbb{X}$) is shown.\label{entropy_diagram}}
\vspace{-10pt} 
\end{figure}

The transfer matrices $\mathbb{T}$ and $\mathbb{X}$ act on $H_B^4$ and can be written as 
\begin{eqnarray}
\mathbb{T}_{i_1i_2i_3i_4}^{j_1j_2j_3j_4}=\sum_{\mu_1,\mu_2 \in \{\pm 1 ,0\}} \bar{A}_{\mu_1}^{i_1j_1}  A_{\mu_1}^{i_2j_2}  \bar{A}_{\mu_2}^{i_3j_3}  A_{\mu_2}^{i_4j_4} \nonumber
\end{eqnarray}
and 
\begin{equation}
\mathbb{X}_{i_1i_2i_3i_4}^{j_1j_2j_3j_4}=
\sum_{\mu_i \in \{\pm 1 ,0\}} \bar{A}_{\mu_1}^{i_1j_1}  A_{\mu_2}^{i_2j_2}  \bar{A}_{\mu_3}^{i_3j_3}  A_{\mu_4}^{i_4j_4}  M_{\mu_1\mu_2\mu_3\mu_4}\nonumber
\end{equation}
where $M\equiv {\rm Tr}[\sigma_{\mu_1}^\dagger \sigma_{\mu_2} \sigma_{\mu_3}^\dagger \sigma_{\mu_4}  ]$. For the case of the \textit{XX} model this can be evaluated explicitly by substituting exact form of $2 \times 2$ $A$ matrices (presented in Sec.\ref{sec:XXmodelResults}) in the expressions of $\mathbb{T},\mathbb{X}$, 
\begin{equation}
\lim_{n_A\to \infty} R_2 (\Delta=0) = - \ln \frac{1+\theta^2/2 + \sqrt{1+\theta^2}}{4}, \text{ where }\theta=\frac{4\epsilon}{4+\epsilon^2}.
\end{equation}

\subsection*{Discussion}
Spin current as a function of $\epsilon$ shows a peak value at $\epsilon=2$ (Fig.~\ref{fig:delta0summary}). The variance of block magnetization as well as entropy per site also shows a minimum at the same point. The domain wall count and staggered magnetization shows a maximum at this point.
The $\epsilon$ dependencies of the means, variances, distributions, rate functions, correlations, and the entropy found in the previous sections occur through combination $2\epsilon/(4+\epsilon^2)$ which is also the expectation value of spin current. The current is same at two different bath couplings $\epsilon$ and $4/\epsilon$. The physical quantities mentioned in the previous sections are the same at these two values of the couplings. This suggests that the state of the system deep in the bulk is determined by the current.

The eigenstates of the current operator on a bond are given by $|\uparrow \uparrow \rangle$, $|\downarrow \downarrow \rangle$, $|\downarrow \uparrow \rangle\pm \imath |\uparrow \downarrow \rangle$ wherein the first two has eigenvalue $0$ while the last two have eigenvalues $\pm J$ i.e they carry spin current in opposite directions. The maximum value of the NESS current injected through the chain using the boundary coupling is however only $\langle J^s\rangle=J/2$ which is less than the largest eigenvalue of the current.

The two site density matrix of the state can be explicitly computed using the two site correlators. We find that the two site density matrix is diagonal in the eigenbasis of the current. Population of the space of fixed current eigenvalues are ($J_s=0$ subspace is 2 dimensional)
\begin{eqnarray}
p({J^s=0})&=\frac{1-\langle J^s\rangle^2}{2},\nonumber\\
p({J^s=\pm J})&=\frac{1+\langle J^s\rangle^2}{4}\pm \frac{\langle J^s\rangle}{2}.\nonumber
\end{eqnarray}
It can be checked that this is the maximum entropy two-site state with fixed $\langle J^s\rangle$.

The states with higher current have higher populations of the two current carrying eigenstates. These two-sites states have antiferromagnetic correlations. This current induced short range antiferromagnetic correlation provides a qualitative picture for the higher domain wall count, broader staggered magnetization distribution and narrow magnetization distribution when the current is higher.

\section{Results for \texorpdfstring{$\Delta\neq 0$}{Delta not 0}\label{sec:XXZmodelResults}}

\subsection{Symmetry transformations}
\label{Symmetry considerations}
We can infer several features of the NESS from the transformation properties of $\mathcal{L}$ and the uniqueness of the NESS. We begin by considering some simple transformations of the Liouvillian.

The unitary transformation $u$ that rotates the spins on alternate sites $1,3,\dots $ by
$\pi$ around the $z$ axis maps the Hamiltonian $H(J,\Delta)$  to $H(-J,-\Delta)$. The jump operators ($\sim\sigma^{\pm}$) may pick up a sign but the dissipators $\mathcal{D}_{1,n}$, being quadratic in the jump operators, are invariant. So $\mathcal{L}(J,\Delta)$ transforms to $\mathcal{L}(-J,-\Delta)$ under $u$. From uniqueness of the NESS we conclude that steady state at parameters $-J,-\Delta$ can be obtained from those at $J,\Delta$ by \[\rho_{\infty}(-J,-\Delta)=u\rho_{\infty}(J,\Delta)u^\dagger.\]

From the definition of $\mathcal{L}$ in Eq.~\ref{Lindblad_equation_def} it is easy to check that $(\mathcal{L}_{J,\Delta}[\rho])^*=\mathcal{L}_{-J,\Delta}[\rho^*]$ in the $z$ basis. This implies that 
\[ \rho_\infty{(-J,\Delta)} = \rho^*_\infty {(J,\Delta)}.\]

Last, we consider unitary transformation $U=\mathcal{I} u_x$ where $\mathcal{I}$ is spatial inversion and $u_x$ is a $\pi$ rotation of all spins about the $x$ axis (which changes the sign of $\sigma_z$ and $\sigma_y$) . The Liouvillian is invariant under this transformation as $u_x$ keeps $H$ invariant but interchanges the $\sigma^{+}$ and $\sigma^-$ thereby interchanging the left and right jump operators; the latter is undone by $\mathcal{I}$. Assuming uniqueness of $\rho_\infty$, it is also invariant under $U$.

The first two results imply that spin profile $\langle S_z \rangle$ is invariant under changes in signs of $J$ and/or $\Delta$. 
For a block of spins in the center of the chain, the last result implies that the generating function $\mathcal{G}^z(\lambda)$ in a block at the center of the chain is an even function of $\lambda$. This implies that the distribution of magnetization and the rate function $W(m=M_z/l)$ are even functions. 

The spin current is given by $J^S=\imath J(\sigma_j^+\sigma_{j+1}^- - \sigma_j^-\sigma_{j+1}^+)$. The first two transformation properties of the NESS imply that $J^S$ is even under change of sign of $J$ and/or $\Delta$.
Similarly, the first two properties imply that expectation value of the energy density $h$ (Eq.~\ref{def_H_XXZ}) changes sign when sign of $J$ or $\Delta$ is changed. They also imply that $x$-magnetization and staggered magnetization interchange roles under change of sign of $\Delta$.

It can also be shown that the energy current $J_E$ at the center of the chain changes sign under the symmetry $U$ of the density matrix, so it has zero expectation value throughout the chain.

\begin{figure}
	\includegraphics{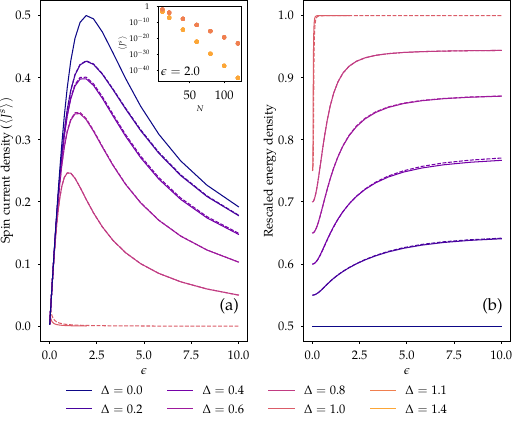}
	\caption{(a) Expectation value of spin current as a function of $\epsilon$ at different $\Delta$. The solid lines and dotted lines are for system sizes $N=100,60$ respectively. (inset) Variation of current in the insulating regime ($\Delta>1$) with system size $N$ for two systems at $\Delta=1.1,1.4$ at $\epsilon=2$. The scale is semi-log i.e. demonstrates $\langle J^s \rangle \propto e^{-cN}$ where $c$ depends on $\Delta$. (b) Rescaled energy density at the center as a function of $\epsilon$. y-axis shows $(h^{NESS}_{j,j+1}-h^{gs}_{j,j+1})/B$ where $h^{gs}$ is the ground state of the bond Hamiltonian and $B$ is the bandwidth of the energy spectrum. $j$ is taken to be the center of the chain for $\Delta<1$ and $j \sim N/4$ for $\Delta>1$.
\label{fig:mean_spin_current_energy_density_exact} }
\end{figure}

\subsection{Expectation value of spin current and energy}
Like the case of $\Delta=0$, spin current as a function of $\epsilon$ has a maximum at some $\epsilon_{\rm max}$ (Fig.~\ref{fig:mean_spin_current_energy_density_exact}~(a)).  
For fixed $\epsilon$ current decreases with $\Delta$. At small $\epsilon$, the current is $\epsilon/2$, independent of $\Delta$. Current is system size $N$ independent for $\Delta<1$, decays exponentially with $N$ for $\Delta>1$ (Fig.~\ref{fig:mean_spin_current_energy_density_exact}~(a) inset). At $\Delta=1$, the current decays as $1/N^2$~\cite{prosen2011exact}. At large $\epsilon$, the current decays as $1/\epsilon$ at $\Delta=0$ (Eq.~\ref{eq:currentExpect}) and $\Delta=1$. 
\begin{figure}
\includegraphics[width=\columnwidth]{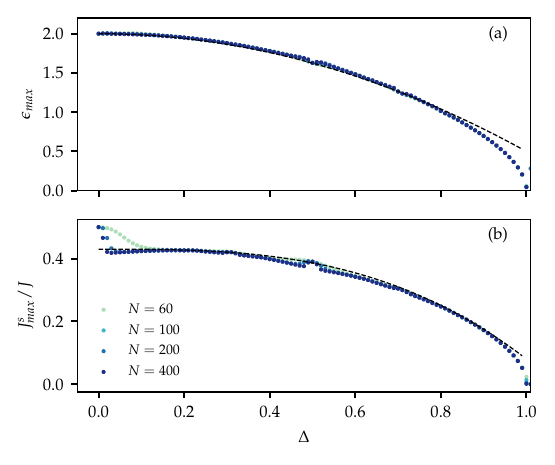}
\caption{
(a) $\epsilon$ corresponding to the maximum of the spin current as a function of $\Delta$. (b) Peak current as a function of $\Delta$. Dotted black lines show the quadratic (panel (a)) and cubic (panel (b)) fits. Different colors denote different system sizes. Black dotted lines are quadratic(cubic) fit to data in panel a(b). (as discussed in main text)
\label{fig:maxCurrent}
}
\end{figure}

The coupling at peak current is $\epsilon_{\rm max}=2$ at $\Delta=0$ and decreases to $\epsilon_{\rm max}=0$ as $\Delta$ increases to $1$ (Fig.~\ref{fig:maxCurrent}(a)). We empirically find this to be closely approximated by $\epsilon_{\rm max}\approx 2-3\Delta^2/2$. The peak current shows a discontinuous jump from ${\rm{max}}[J^s]=J/2$ to a smaller value as $\Delta$ changes from $0$. The discontinuity appears to be smoothened by the finite size effect. For $\Delta>0$, we empirically find that the peak current is well approximated by ${\rm{max}}[J^s]=0.43-0.35\Delta^3$ till $\Delta$ close to $1$. 

Energy current is exactly $0$ due to the symmetries discussed in the previous section. Energy density is a constant deep in the bulk for $0<\Delta<1$. This bulk value increases monotonically with $\epsilon$ and $\Delta$ as shown in Fig.~\ref{fig:mean_spin_current_energy_density_exact}(b). 

\begin{figure}
\includegraphics[width=0.8\columnwidth]{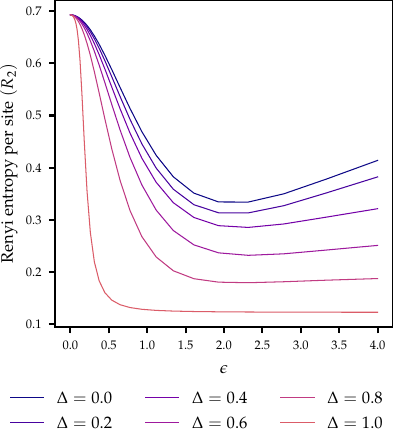}
\caption{ Bipartite Renyi Entropy per site ($R_2$) as a function of $\epsilon$ for different values of $\Delta$. $N=20$ and $N_A=10$.   \label{fig:EntropyDelta}}
\end{figure}

\subsection{Entropy\label{sec:finitedeltaEntropy}}
Numerically obtained values of entropy per site in a bipartitite system with equal halves is shown in Fig.~\ref{fig:EntropyDelta}. The entropy is $\ln 2$ for $\epsilon=0$ and $\epsilon=\infty$ representing an infinite temperature state at $\Delta=0$. This appears to hold for small $\Delta$ as well in finite size calculations. The bipartite Renyi entropy ($R_2$) decreases with $\epsilon$ initially till some $\epsilon_{\rm min}^{\rm entropy}(\Delta)$ and then increases at larger $\epsilon$. Entropy decrease with $\Delta$ until $\Delta=1$, after which the spins are nearly maximally polarized and have $R_2\approx 0$. 

Qualitative features for the von Neumann entropy of small subsystems of size $4$ showed similar features. Entropy is lower at the ends and higher in the bulk. Unlike the case of $\Delta=0$, the minimum of the entropy as a function of $\epsilon$ does not coincide with current maximum.

\subsection{Spin correlations}
\begin{figure}
	\includegraphics[width=\columnwidth]{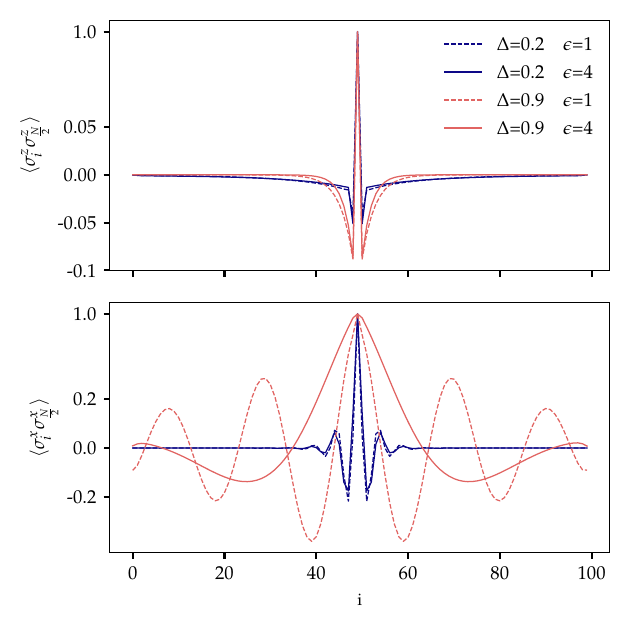}
	\caption{Representative examples of the equal time correlation of the spin at the center of the chain with other spins as a function of position($i$) . $\langle \sigma^z_i \sigma^z_{\frac{N}{2}} \rangle$ (panel (a)) and $\langle \sigma^x_i \sigma^x_{\frac{N}{2}} \rangle$ (panel (b)) for different values of $\Delta$ and $\epsilon$. Solid and dashed lines denote $\epsilon=4$ and $1$ respectively.   \label{fig:corr_data}}
\end{figure}
Figure~\ref{fig:corr_data}(a) shows the connected correlator between the $z$ components of the spins at the center and other sites. Qualitative properties of the $z$-correlations remain the same as that in $\Delta=0$, adjacent sites are anti-correlated but correlations at larger distance remain weak even at larger $\Delta$ and $\epsilon$. 

Figure~\ref{fig:corr_data}(b) shows the representative examples of correlations between the $x$-components of the spins. $\langle \sigma^x_i \sigma^x_{\frac{N}{2}} \rangle$ is short ranged but stronger than $\langle \sigma^z_i \sigma^z_{\frac{N}{2}} \rangle$. It is oscillatory with an envelop rapidly decaying with distance. $x$-correlations appear stronger at larger $\Delta$ and $\epsilon$. In the limit of large $N$, correlation lengths show strong discontinuous $\Delta$ dependence, and are consistent with the eigenvalues of the transfer matrix. The correlations are primarily short ranged. However, asymptotic tails containing small amplitude correlations that decay slowly with distance are present.

\subsection{Scaled cumulant generating function and rate function for magnetization \label{SCGF}}
In this section, we present the SCGF and rate function (also referred to as large deviation function(LDF)) for the magnetization density $m=M_l^z/l$. This characterizes the magnetization distribution in a block of $l$ sites in the middle of a chain of length $N=2n+l$ where $n$ denotes the chain length outside the block of interest on both sides. The SCGF $\phi_{M_l^z}(\lambda)$ is calculated as $\frac{1}{l} \ln \langle e^{\lambda M_l^z} \rangle$ for real valued $\lambda$. Unlike the $\Delta=0$ case, the SCGF cannot be analytically calculated. However, we can compute the SCGF in a numerically exact way using the transfer matrix approach (Sec.~\ref{sec:XXmodelResults}).  The large deviation function can be calculated in principle by Legendre transformation after taking $N\to \infty$ (this is equivalent to $n \to \infty$) followed by $ l\to \infty$ limit on SCGF. We choose $\Delta$ values such that $\omega=\arccos{\Delta}$ is a rational multiple of $\pi$ i.e $\omega=\frac{p}{q} \pi$ where $p,q$ are co-prime integers. For these cases, we can access the asymptotic-$N,l$ behavior of the SCGF. Accessing asymptotic-$N$ limit is not possible for generic $\Delta$ as the transfer matrices have infinite dimesion in those cases when $N \to \infty$. We focus on the $\omega=\frac{p}{q} \pi$ cases in the rest of this section.

The probability distribution $P(m)$ of local observables like $M^z_l$ can also be obtained directly as the inverse Fourier transform of the imaginary field SCGF $\phi_{M^z_l}(\imath \lambda)=\frac{1}{l} \ln \langle e^{\imath \lambda M_l^z} \rangle$ for finite systems. The rate function can then be computed as $-\frac{1}{l} \ln \frac{P(m)}{P(0)}$. The probability of large  deviations from mean value of magnetization becomes exponentially suppressed for large $l$,  requiring numerically expensive high-precision calculations. 
We present a comparison between this approach and the rate function estimated from Legendre transform in Appendix~\ref{Appendix: compare rate functions} demonstrating the equivalence of both approaches in systems as large as $l=280$ and $N=400$.

In the remainder of this section, we exclusively present the results for the rate function estimates obtained from the numerical Legendre transform of SCGF. We present the data for $\epsilon = 0.1,1.0$ and $10.0$ as demonstrative cases. 

The finite-size SCGF is calculated for the steady state as follows,
\begin{align}
\label{scgf expression}
\phi_{M^z_l}(\lambda,l,N)=\frac{1}{l} \ln  {\rm Tr}(\rho_{\infty} e^{\lambda \sum_{k=1}^{l} \sigma^k_z} )
\end{align}
which can be evaluated using the transfer matrix approaches in Sec.~\ref{Expectation value from transfer matrix}.  It can be inferred by looking at the elements of transfer matrix (See Appendix \ref{Matrix elements of A}) and MPO representation of the generating function that the corresponding matrices get truncated to the dimension $ q$ for $\Delta = \cos(p \pi /q)$. This allows us to get asymptotic behavior of SCGF and rate function in a numerically exact way. The transfer matrix, $T$ has a spectral decomposition $T=\sum\limits_{i=1}^q t_i \mathcal{T}_{i,R} \mathcal{T}^T_{i,L}  $ where $t_i,\mathcal{T}_{i,R},\mathcal{T}_{i,L}$ are $i$-th eigenvalue, corresponding right and left eigenvectors of $T$ respectively.

To compute SCGF, we need the transfer matrix corresponding to $e^{\lambda M_l^z}$ as well. We represent it as $U$ which has the following expression. 
\begin{align}
    U(\lambda)= \sum_{r=0}^{\infty} {e^{-\lambda}|a_r^+|^2}  |r \rangle \langle r+1| +  e^{\lambda}|a_r^-|^2 |r+1 \rangle \langle r|+ \nonumber \\ 
      {2\cosh(\lambda) |a_r^0|^2 }  |r \rangle \langle r|
\end{align}
Equation~\ref{scgf expression} can be expanded as,
\begin{align}
\label{scgf expression_expanded}
\phi_{M^z_l}(\lambda,l,N)=\frac{1}{l} \ln \frac{\langle 0 |T^n U^l T^n |0\rangle}{\langle 0 |T^{2n+l} |0\rangle}
\end{align}
where we have assumed the system has $N=2n+l$ sites in total and observables are calculated in a  block of length $l$ in the middle of the chain. In the $n \to \infty $ limit with finite $l$, this further simplifies to
\begin{align}
\label{scgf_expression_expanded_asymptotic_n}
\phi_{M^z_l}(\lambda,l,\infty)&=\lim_{N \to \infty }\phi_{M^z_l}(\lambda,l,N) \nonumber \\
&=\frac{1}{l} \ln \frac{\mathcal{T}^T_{1,L}  U^l \mathcal{T}_{1,R} }{\mathcal{T}^T_{1,L} \mathcal{T}_{1,R} } \left( \frac{1}{t_1}\right )^l.
\end{align}
 $\mathcal{T}_{1,R}, \mathcal{T}^T_{1,L} $ and $ t_1 $ are dominant right and left eigenvectors and corresponding eigenvalue of $T$ respectively. By considering a similar spectral decomposition of $U=\sum\limits_{i=1}^q u_i \mathcal{U}_{i,R} \mathcal{U}^T_{i,L}$, we can obtain the asymptotic $N,l$ limit of SCGF,$\Phi_{M^z_l}(\lambda)$. We have numerically verified that for all the cases studied here, the spectral gap between two dominant eigenvalues of $T$ and $U$ is finite which justifies the following expressions. The spectral gap goes down as $\Delta \to 1$. (see Appendix~\ref{Appendix: dominant eigenvalues}),

\begin{figure}
	\includegraphics[width=\columnwidth]{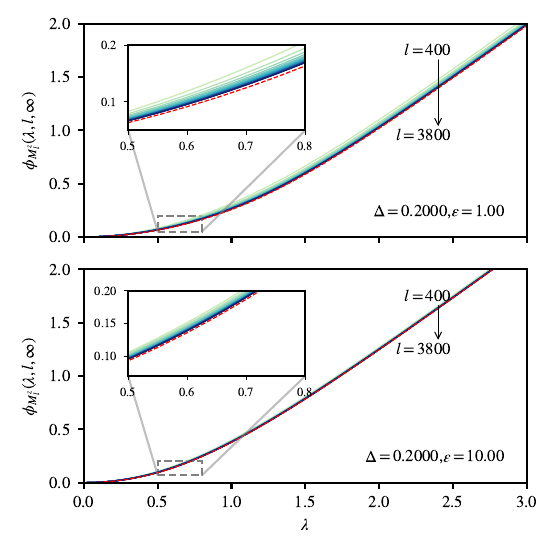}
	\caption{The SCGF numerically estimated for finite $l$ and $N \to \infty$ for $\Delta=\cos(17\pi/39) \approx 0.2$ and $\epsilon \in [1.0,10.0]$. Insets show the zoomed in regions. The color grading goes from light green ($l=400$)) to blue ($l=3800$) in step of 400. The red dotted line indicates SCGF at $l \to \infty$ obtained using Eq.~\ref{scgf_expression_expanded_asymptotic_nl}.~\label{fig:scgf_finite_size}}
\end{figure}

\begin{align}
\label{scgf_expression_expanded_asymptotic_nl}
\Phi_{M^z_l}(\lambda)&=\phi_{M^z_l}(\lambda,\infty,\infty)\nonumber \\ 
&=\lim_{l \to \infty } \phi_{M^z_l}(\lambda,l,\infty) \nonumber \\
&=\lim_{l \to \infty }\frac{1}{l} \ln \frac{(\mathcal{T}^T_{1,L} \mathcal{U}_{1,R}) (\mathcal{U}^T_{1,L} \mathcal{T}_{1,R}) }{(\mathcal{T}^T_{1,L} \mathcal{T}_{1,R}) (\mathcal{U}^T_{1,L} \mathcal{U}_{1,R})}\left(\frac{u_1}{t_1}\right )^l \nonumber \\
&= \ln \frac{u_1}{t_1}
\end{align}
Expanding $\Phi_{M^z_l}(\lambda)$ up to second order in $\lambda$ we can obtain the asymptotic-$N,l$ variance. This will be discussed in the next subsection.
Subsequently, leading order finite-$l$ correction to 
$\phi_{M^z_l}(\lambda,\infty,\infty)$ can be estimated by expanding Eq.~\ref{scgf_expression_expanded_asymptotic_n},
\begin{align}
\label{leading_correction_scgf}
&\phi_{M^z_l}(\lambda,l,\infty) -\Phi_{M^z_l}(\lambda) \nonumber \nonumber \\ 
&= \frac{1}{l} \ln \left(1+\alpha \left(\frac{u_2}{u_1}\right)^l \right) +\text{subleading terms},
\end{align}

where $\alpha=\frac{(\mathcal{T}^T_{1,L} \mathcal{U}_{2,R}) (\mathcal{U}^T_{2,L} \mathcal{T}_{1,R}) }{(\mathcal{T}^T_{1,L} \mathcal{U}_{1,R}) (\mathcal{U}^T_{1,L} \mathcal{T}_{1,R}) }$. This suggests that SCGF approaches the asymptotic-$l$ value $\Phi_{M^z_l}(\lambda)$ as $\frac{C_1}{l}e^{-C_2 l}$ where $C_1$ and $C_2$ depends on $\Delta,\epsilon$ and counting field $\lambda$ i.e $l$ dependence of SCGF is asymptotically exponential.

We present the SCGF obtained for finite-$l$ and its asymptotic-$l$ behavior in Fig.~\ref{fig:scgf_finite_size} for $\Delta=\cos(17\pi/39) \approx 0.2$ and $N \to \infty$. The finite-$l$ SCGF is obtained using Eq.~\ref{scgf_expression_expanded_asymptotic_n}. 
The finite-size effect is largest at intermediate $\epsilon$.

\begin{figure}
	\includegraphics[width=\columnwidth]{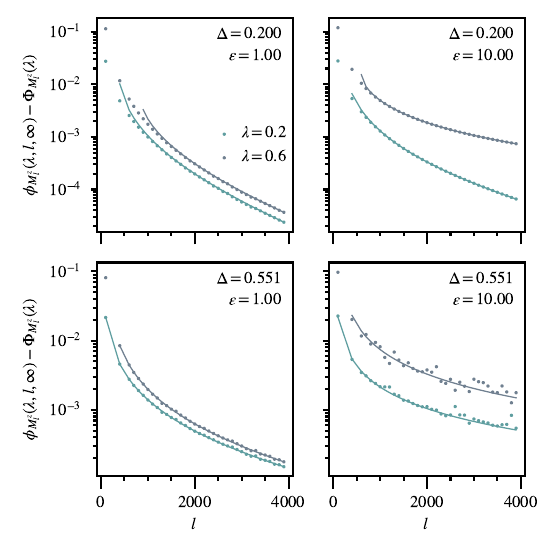}
	\caption{The first order correction to asymptotic SCGF as a function of blocksize $l$ at $N \to \infty$. We represent the data for $\Delta=\cos(17\pi/39) \approx 0.2 $ (top panel) and $\Delta=\cos(11 \pi /35) \approx 0.551$ (bottom panel) . The dots denote the difference of finite-$l$ SCGF from its asymptotic limit. The solid lines indicate the leading order correction obtained from Eq.~\ref{leading_correction_scgf}. The data shows two values of counting field ,$\lambda=0.2$ and $0.6$.
   \label{fig:leading_order_corection_scgf}}
\end{figure}

We show the convergence of SCGF to its asymptotic-$l$ value in Fig.~\ref{fig:leading_order_corection_scgf} for two values of counting field $\lambda$ at multiple $\epsilon,\Delta$. The deviation from asymptotic limit of SCGF due to finite $l$ is almost entirely captured by the first order terms in Eq.~\ref{leading_correction_scgf} (shown using solid lines in Fig.~\ref{fig:leading_order_corection_scgf}). The decay constant $C_2$ depends on the gap between the lowest two eigenvalues $u_2,u_1$. For $\Delta$ of the form $\cos \frac{p}{q}\pi$, the gap is shown as a function of $\Delta$ in Fig.~\ref{fig:ratio_eigenvalues_V} in Appendix~\ref{Appendix: dominant eigenvalues}. The gap decreases with $p$ and $q$. This implies that even if $p/q$ and $p'/q'$ are arbitrarily close to each other, if $p',q'$ are larger than $p,q$, then their gaps will differ. Thus the gap is a highly discontinuous function of $\Delta$. 

From the SCGF, we can then obtain a finite-size estimate of the rate function $I(m,l,N)$ using a Legendre transform of the SCGF (assuming G\"artner-Ellis theorem\cite{gartner,ellis1984large}),
\begin{align}
    I(m,l,N)=\sup_{\lambda \in \mathcal{R}}\{\lambda m - \phi_{M^z_l}(\lambda,l,N)\}
\end{align}

\begin{figure}
	\includegraphics[width=\columnwidth]{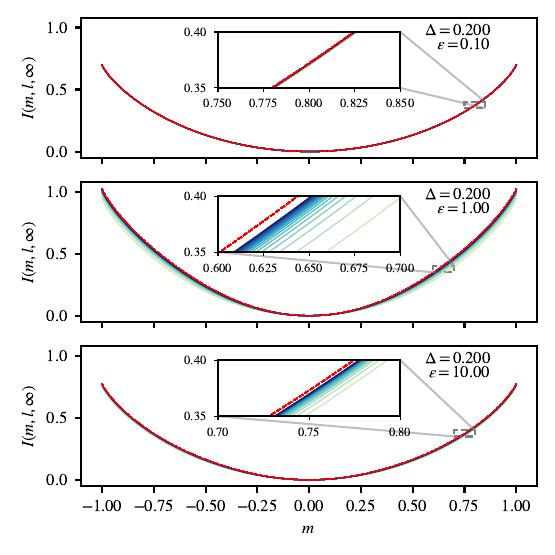}
	\caption{The rate function corresponding to the distribution of the average magnetization $m=M_z/l=\sum \sigma_z/l$ in a block of size $l$ for NESS at $\Delta=\cos(17\pi/39) \approx 0.2$ and $\epsilon=0.1,1.0$ and $10.0$ at $N \to \infty$. Different lines represent different $l$. The color grading goes from light green ($l=400$)) to blue ($l=3800$) in step of 400. The red dotted line is the asymptotic large-$l$ rate function obtained from Legendre transformation of Eq.~\ref{scgf_expression_expanded_asymptotic_nl}
. (Inset) The zoomed in view of the main figure at extreme $m$.\label{fig:rate_function_l_asymptotic}}
\end{figure}

Figure~\ref{fig:rate_function_l_asymptotic} shows the convergence of finite-$l$ estimate of the rate function for average magnetization $m=M^z_l/l$ in a block of size $l (400 \leq l \leq 3800) $ for multiple $\epsilon$ values in the $n \to \infty$ limit. The finite-size effect in the estimate of SCGF is expected to show up in rate function estimate as well. This can be clearly seen in the inset. Note that the finite-$l$ effect is minimal at small $\epsilon$ regime, grows in intermediate $\epsilon$ and is suppressed in large $\epsilon$. 

Finally, to complete the discussion, we study the finite-$N$ effect in rate function and its scaling properties in Appendix~\ref{Finite-N-rate-function}.


\subsection{Properties of magnetization distribution }
\begin{figure}
	\includegraphics[width=\columnwidth]{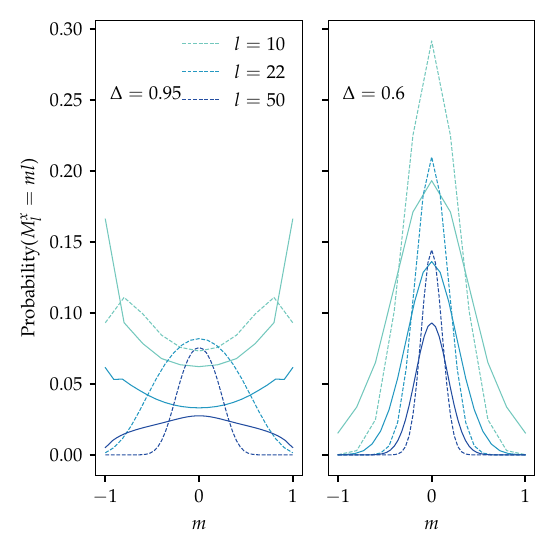}
	\caption{Distribution of $M^x_l$. (a) $\Delta=0.95$ for  $\epsilon=1$(dashed lines) and $4$ (solid lines). (b) $\Delta=0.6$ for  $\epsilon=1$(dashed lines) and $4$ (solid lines). Legend shows the blocksizes ($l$) studied. \label{fig:Sx_distribution}}
\end{figure}
In this section, we look at the distribution of transverse magnetization in the block. This calculation cannot be performed in large systems as the transfer matrices cannot be restricted to the charge $Q=0$ block of the bond Hilbert space. Nevertheless explicit calculation can be performed in smaller systems.

Figure~\ref{fig:Sx_distribution} shows the distribution of $M^x_l = \sum\limits_{i=1}^{l} \sigma^x_i$ in a block of size $l$ in the middle of a chain of $N=100$ sites for different values of $\epsilon$ and $\Delta$. At large values of $\Delta>0.9$ the distribution shows a double peak structure in small block sizes, which merges into a single peak in larger blocks. This suggests a short range ferromagnetic ordering at large $\Delta$ for $\sigma^x$. The ordering is stronger - they persist to larger blocks - as $\epsilon$ increases. Similar short range ordering was observed in the ground state of \textit{XXZ} chain at large $\Delta$~\cite{collura2017full}. At $\Delta<0.9$ no double peak structure was observed.

When $\Delta$ is negative, from symmetry considerations (See Sec:\ref{Symmetry considerations}), we see that this short range $x$ order is antiferromagnetic in nature.

\begin{figure}
	\includegraphics[width=0.8\columnwidth]{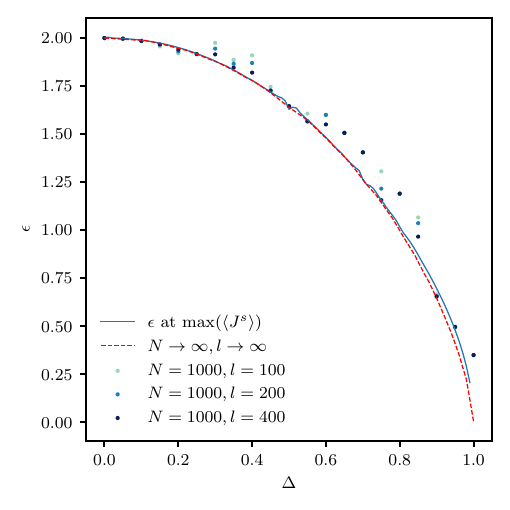}
	\caption{ Comparison of $\epsilon$ values where $\langle J^s \rangle$ attains maximum (solid blue line) and var($M^z_l$) attains minimum. The solid blue line represents the $\epsilon$ where $\langle J^s \rangle $ is maximum. The dashed red line shows the $\epsilon$ where var($M^z_l$) is minimum at asymptotic-$l,N$ limit. The dots are from the variance obtained in finite size calculation. 
	\label{fig:min_current_var_Mz}}
\end{figure}

Finally, we discuss the trend of variance of $M^z_l$ with $\langle J^s \rangle$. It can be expressed as follows,
\begin{align}
\label{var_expression}
\rm{var}(\mathit{M}^z_l)=\rm{Tr}(\rho_{\infty }(\mathit{M}^z_l)^2)-(\rm{Tr}(\rho_{\infty }\mathit{M}^z_l))^2
\end{align}
The second term (i,e. mean $z$-magnetization) in the expression for variance is $0$ for $\Delta<1$. In this case as well, we can study the asymptotic-$N,l$ behavior at $\omega=p \pi /q $ points i.e rational multiple of $\pi$. Equation~\ref{var_expression} evaluates to the following in asymptotic-$N,l$ limit,
\begin{align}
\label{asymptotic var}
\lim_{l \to \infty}\lim_{N \to \infty}\rm{var}
(M^z_l)=\frac{(\mathcal{T}^T_{1,L} \mathit{T}_{\text{diag}} \mathcal{T}_{1,R})
}{(\mathcal{T}^T_{1,L} \mathcal{T}_{1,R}) \mathit{t_1}}
\end{align}
where $T_{diag}$ denotes the diagonal part of of transfer matrix, $T$. This can be obtained from expansion of Eq.~\ref{scgf_expression_expanded_asymptotic_nl} and keeping the $O(\lambda^2)$ term. We present data for finite-$N,l$ by taking numerical second derivative of Eq.~\ref{scgf expression_expanded} with $\lambda$. The data is presented in Fig.~\ref{fig:min_current_var_Mz}. We also present the asymptotic-$N,l$ behavior of variance with the dashed red line (Eq.~\ref{asymptotic var}).  

We observe that the $\epsilon$ at minimum of var($M^z_l$) overlaps with $\epsilon$ where maximum of $\langle J^s \rangle $ is attained at asymptotic limit. The $\epsilon$ at minimum of var($M^z_l$) deviates from the $\epsilon$ at max($\langle J^s \rangle $) for finite size calculations. This observation further supports the effect of correlation induced by spin current in the bulk for finite $\Delta$. 

\section{Conclusion\label{sec:conclusion}}

In this work, we have characterized the current carrying NESS of an \textit{XXZ} model with maximally polarizing Lindblad couplings at the boundary by calculating the spin correlations and full distribution of the spin densities. The calculation of the full distribution function which in general requires accurate information about the steady state is made possible by the exact solution of the NESS in a MPS form~\cite{prosen2011exact}. 
The state carries a current that can be tuned via the strength of the coupling to the bath $\epsilon$. The current initially increases and then decreases as $\epsilon$ is increased from 0. Bulk properties such as the correlations, distribution of observables and the entropy in the steady state are determined by both the Hamiltonian parameters and the current in the system.
Though the MPS form does not yield closed form expressions in general, calculations of correlations and distribution functions are reduced to multiplications of transfer matrices of dimensions at most linear in the system size enabling numerically exact calculations.

The NESS in the $\Delta=0$ case corresponding to the \textit{XX} model has an especially simple $2\times 2$ transfer matrix; and we can get exact closed form expressions for various quantities such as spin-spin correlations in the XY plane and along $z$ directions, entropy per site etc. The scaled cumulant generating functions of magnetization, staggered magnetization and domain wall densities in the $z$-direction can also be calculated exactly allowing explicit demonstration of the large deviation principle. These also allow numerically exact calculation of rate functions using numerical Legendre transforms. The rate functions do not show any indication of a singularity as a function of the current or $\epsilon$. 
At least at the level of the two site density matrix, the state is described by a maximal entropy ensemble with a constraint on its spin current expectation value. The various physical quantities that we studied, namely the entropy density, spin correlators and distributions (of $z$-magnetizations, staggered magnetizations, domain wall count) depends on the $\epsilon$ only through the current expectation value $J\propto=\frac{2\epsilon}{4+\epsilon^2}$. 
The variances of the distributions show an extremum when the current is maximized as a function of the boundary coupling constant. In particular, the $z$-magnetization shows a minimum variance when the current is maximized. This can be qualitatively understood from the increased population of the antiferromagnetic two site states as the current is increased. This results in a reduction of the probability of occurrence of large-magnetization states, and therefore, in a narrower $z$-magnetization distribution.

All magnetization distribution functions in the \textit{XX} model are qualitatively similar to that of the infinite temperature ensemble as a result of the short range spin correlations in the NESS. The current work explored only the case where spin-chemical-potential $\mu$ at the ends are maximal ($\pm 1$) and when the dynamics has an exact $U(1)$ rotation symmetry about the z-axis. Using a fermionic representation of the spins, different from what has been used here, the NESS in the \textit{XY} model can be studied at $|\mu|\neq 1$. These are known to exhibit unusual nonequilibrium phase transitions~\cite{vznidarivc2011solvable,PhysRevLett.101.105701}. It will be interesting to see if and how these phase transitions impact the full distributions. We leave this for future investigations. 

Using the MPS solution, we could obtain numerically exact calculations for systems at finite $\Delta$ as well. We obtain the rate function at asymptotic limit for cases where $\Delta=\cos(\omega)$ with $\omega=p \pi/q$ and present a detailed analysis on convergence with blocksize. The results for these satisfy the large deviation principle for local observables in an interacting strongly driven quantum system. We find that finite block-size corrections in the rate function estimation are higher at the intermediate $\epsilon$ when the current is maximal. We find that narrowing of the $z$- magnetization distribution when the current is maximized happens in the case of $\Delta>0$ as well. Surprisingly, the exponential decay rate of approach to the asymptotic form of the SCGF is a strongly discontinuous function $\Delta$, with the decay rate vanishing as $p,q$ increase. This suggests that for irrational values of $\omega$, the finite block-size effect will be much stronger, it is unclear from our calculations if these satisfy large deviation principle.

The $x$-magnetization distribution in large systems show qualitatively similar behavior to the $z$-magnetization possibly due, again, to weak long range correlations. However at large $\Delta\gtrapprox 0.9$ and large $\epsilon$, the $x$-distributions show a prominent double peak structure in small blocks indicating short range ordering in the $x$-direction, similar to what was described in existing literature on the ground state of the \textit{XXZ} model~\cite{collura2017full}.

{\acknowledgements}
We thank National Supercomputing Mission (NSM) for providing computing resources of ``PARAM Brahma" at IISER Pune, which is implemented by C-DAC and supported by the Ministry of Electronics and Information Technology (MeitY) and Department of Science and Technology (DST), Government of India.
\bibliography{exactfcs}
\clearpage
\onecolumngrid
\appendix
\pagenumbering{alph}
\section{Solution to divergence condition\label{sec:Asolutions}}
In this Appendix, we find solutions for $S_{N}$ that satisfy Eq.~\ref{eq:divergence_condition} and Eq.~\ref{Eq:SRLrelations}. We first find the algebra of $A$ tensors that solve the conditions and in the subsequent section construct an explicit form of $A$ that solves the algebra.	

First we argue that the commutator $[H,S_N]$ is a combination of Pauli strings that each have $\sigma^{z}$ on exactly one site. It is sufficient to show this for individual terms of the commutator:
\[
\left[h_{j,j+1},S_N\right]=\sum_{\left\{ r_{i}\right\} }\langle0|A_{r_{1}}\dots A_{r_{n}}|0\rangle\sigma^{r_{1}}\sigma^{r_{2}}\dots\sigma^{r_{j-1}}\left[h_{j,j+1},\sigma^{r_{j}}\sigma^{r_{j+1}}\right]\sigma^{r_{j+2}}\dots\sigma^{r_{n}}
\]
We have omitted the tensor product symbols for compactness.
The commutator affects only the sites $j,j+1$. From explicit evaluation, it can be seen that the commutator of any of the terms inside $h_{j,j+1}$ namely $\sigma_{+}^{j}\sigma_{-}^{j+1}$, $\sigma^{-}_{j}\sigma^{+}_{j+1}$ or $\sigma^{z}_{j}\sigma^{z}_{j+1}$ with $\sigma^{r_{j}}\sigma^{r_{j+1}}$ for any $ r_{j},r_{j+1}\in\left\{ \pm,0\right\} $ either vanishes or produces terms that contain exactly one $\sigma^{z}$. Thus $[H,S_N]$ is a linear combination of Pauli strings with $\sigma^{z}$ occurring exactly once in each string. 

Equation~\ref{eq:divergence_condition} demands that among these, only the strings in which $\sigma^{z}$ occur at the ends have finite amplitude, which implies that for all sites other than the two ends i.e. for sites $2\leq j\leq N-1$ the following is true
\[{\rm Tr}\left[\sigma^{a\dagger}_{j-1}\sigma^{z\dagger}_j\sigma^{b\dagger}_{j+1}\left[h_{j-1,j}+h_{j,j+1},S_N\right]\right]=0\text{ for all }a,b\in\left\{ \pm1,0\right\}.\]
This is satisfied if 
\begin{equation}
\sum_{r_{1},r_{2},r_{3}\in \{\pm,0\}}{\rm Tr}\left[\sigma^{a\dagger}\sigma^{z\dagger}\sigma^{b\dagger}\left[h_{12}+h_{23},\sigma^{r_{1}}\sigma^{r_{2}}\sigma^{r_{3}}\right]A_{r_{1}}A_{r_{2}}A_{r_{3}}\right]=0\text{ for all }a,b\in\left\{ \pm1,0\right\}\label{eq:constraintMain}
\end{equation}
These nine equations result in eight constraints of the form
\begin{eqnarray}
&\left[A_{0},A_{\pm}A_{\mp}\right]=0\label{eq:cond1}\\
&\left\{ A_{0},A_{\pm}^{2}\right\} =2\Delta A_{\pm}A_{0}A_{\pm}\label{eq:cond2}\\
&\Delta A_{0}^{2}A_{\pm}-A_{0}A_{\pm}A_{0}=A_{\mp}A_{\pm}^{2}-A_{\pm}A_{\mp}A_{\pm}\label{eq:cond3}\\
&\Delta A_{\pm}A_{0}^{2}-A_{0}A_{\pm}A_{0}=A_{\pm}^{2}A_{\mp}-A_{\pm}A_{\mp}A_{\pm}\label{eq:cond4}
\end{eqnarray}
which are equivalent to the constraints given in Eq. 9 of Ref.~\cite{prosen2011exact}. The first line is obtained from considering $\left(a,b\right)=\left(2,3\right)$ and $\left(3,2\right)$ in Eq.~\ref{eq:constraintMain}, second line from $\left(2,2\right)$ and $\left(3,3\right)$, the third line from $\left(1,3\right)$ and $\left(1,2\right)$, and last line from $\left(3,1\right)$ and $\left(2,1\right)$. These constraints eliminate occurrence of $\sigma^{z}$ on any site $2\leq i<N$.

Now we consider the terms in $[h_{12},S_N]$ that contain a $\sigma_{z}^{1}$. These should be nonvanishing such that the RHS of Eq.~\ref{eq:divergence_condition} is reproduced. $[h_{12},S_N]$ can be explicitly calculated to be 
\[\sqrt{2}\sigma_1^{z}\left[\sigma_2^{-}M+\sigma_2^{+}P+\sigma_2^{0}E\right]\times \sum_{r_{3},r_{4},\dots,r_{N}}A_{r_{3}}A_{r_{4}}\dots A_{r_{N}}|0\rangle\prod_{i=3}^{N}\sigma^{r_{i}}\]
where 
\begin{eqnarray}
&M=\langle0|A_{-}A_{0}-\Delta\langle0|A_{0}A_{-}\nonumber\\
&P=-\langle0|A_{+}A_{0}+\Delta\langle0|A_{0}A_{+}\nonumber\\
&E=\langle0|A_{-}A_{+}-\langle0|A_{+}A_{-}\nonumber
\end{eqnarray}
To satisfy Eq.~\ref{eq:divergence_condition}, the above terms should reproduce $\sigma_z L$. Comparing coefficients of $\sigma_\pm,\sigma_0$ on the second site, we see that the commutator correctly produces the $\sigma_z L$, if  $P=-\imath\epsilon\langle0|A_{+}/\sqrt{2}$,
$M=-\imath\epsilon\langle0|A_{-}/\sqrt{2}$ , and $E=-\imath\epsilon\langle0|A_{0}/\sqrt{2}$ are satisfied.
These are satisfied if the following conditions are imposed on the $A$ operators:
\begin{equation}
\langle0|A_{0}=\langle0|\sqrt{2},\quad\langle0|A_{-}=0,\quad\langle0|A_{+}A_{-}=\imath\epsilon\langle0|\label{eq:boundaryConds}
\end{equation}
To see that these solve the constraint $P=-\imath \epsilon \langle0|A_{+}$, we multiply both sides of Eq.~\ref{eq:cond3} by $\langle0|$ to get
\[
\Delta\langle0|A_{0}^{2}A_{+}-\langle0|A_{0}A_{+}A_{0}=\langle0|A_{-}A_{+}^{2}-\langle0|A_{+}A_{-}A_{+}
\]
This reduces to the desired equation $\sqrt{2}P=-\imath \epsilon\langle0|A_{+}$ if Eq.~\ref{eq:boundaryConds} are satisfied.

Similar analysis on the right edge gives the following boundary conditions 
\begin{equation}
A_{0}=\sqrt{2}|0\rangle,\quad A_{+}|0\rangle=0,\quad A_{+}A_{-}|0\rangle=\imath\epsilon |0\rangle\label{eq:boundaryCondsR}
\end{equation}
Eq.~\ref{eq:boundaryConds}, Eq.~\ref{eq:boundaryCondsR} also imply Eq.~\ref{Eq:SRLrelations}.

\subsection{Finding the matrix elements of \texorpdfstring{$A$}{A} \label{Matrix elements of A}}

Now we construct explicit matrix form of the algebra given in Eq.~\ref{eq:cond1}-\ref{eq:cond4} and the boundary conditions Eq.~\ref{eq:boundaryConds},\ref{eq:boundaryCondsR} assuming the tridiagonal structure of the solutions shown in Eq.~\ref{eq:Amatrices} of the main text.
We do this by writing Eq.~\ref{eq:cond1}-\ref{eq:cond4} and Eq.~\ref{eq:boundaryConds},\ref{eq:boundaryCondsR} in terms of $a_{\pm}$ and $a_{0}$. This gives the following conditions on $a$'s
\begin{eqnarray}
    &a_{r}^{0}+a_{r+2}^{0}  =2\Delta a_{r+1}^{0}\text{ for }r\geq0\label{eq:a0}\\
    &2(\Delta a_{0}^{0}-a_{1}^{0})a_{0}^{0}=-a_{0}^{+}a_{0}^{-}\label{eq:a0bcs}\\
    &2(\Delta a_{r}^{0}-a_{r+1}^{0})a_{r}^{0}=a_{r-1}^{+}a_{r-1}^{-}-a_{r}^{+}a_{r}^{-}\text{ for }r>0\label{eq:eq:other}
\end{eqnarray}
with the boundary conditions 
\begin{equation}
    a_{0}^{0}=1,\quad a_{0}^{-}=1\quad a_{0}^{+}=\imath\epsilon
\end{equation}
Equations ~\ref{eq:a0} and ~\ref{eq:a0bcs} along with the boundary conditions have the solution (for $r\geq 0$)
\begin{equation}
a_r^0 = \cos (r\omega) + \imath \epsilon \frac{\sin (r\omega)}{2\sin \omega}\label{eq:a0solution}
\end{equation}
where we defined $\omega$ by $\cos\omega = \Delta$. The recursion relation in Eq.~\ref{eq:eq:other} allows us to obtain 
\begin{equation}
a_r^+a_r^- = \imath \epsilon - \sum_{s=0}^r 2 (\Delta a_{s}^{0}-a_{s+1}^{0})a_{s}^{0}
\end{equation}
The summation can be performed exactly after plugging in the solution for $a^{0}_{s}$ to get
\begin{equation}
a_r^+ a_r^- = -\left[1+\frac{\epsilon^2}{4\sin^2\omega}\right] \sin (r\omega) \sin (r\omega+\omega) + \frac{\imath \epsilon}{\sin \omega} \cos (r\omega)\, \sin (r\omega+\omega)
\end{equation}
Solution to these can be chosen to be (for $r\geq 1$)
\begin{equation}
a_r^+ = \frac{\sin (r\omega+\omega)}{\sin \omega};\quad a^-_r = -\left[1+\frac{\epsilon^2}{4\sin^2\omega}\right] \sin (r\omega) \sin\omega + {\imath \epsilon} \cos (r\omega) 
\label{eq:AppendixAmatrixElements}
\end{equation}
From the expressions for $a_r^+$ and $a_r^-$, we can readily observe that $a_r^+=0$ or $a_r^+=0$ at $r=q$ when $\omega=p/q$ i.e a rational fraction (depending on $q$ being even or odd). This makes it possible to truncate the corresponding transfer matrix to a $q \times q $ block as the the vanishing of $a_q^+$ prohibits transition from state $q$ to $q+1 \in H_B \times H_B$. We utilize this property in Sec.~\ref{SCGF} in main text.

An alternative solution for coefficients in $A_\pm$ is given in Ref.~\cite{prosen2011exact}, we have checked extensively that the two produce identical results.

\section{Rate function from full distribution\label{Appendix: compare rate functions}}
\begin{figure}[H]
    \centering
    \includegraphics[width=0.5\textwidth]{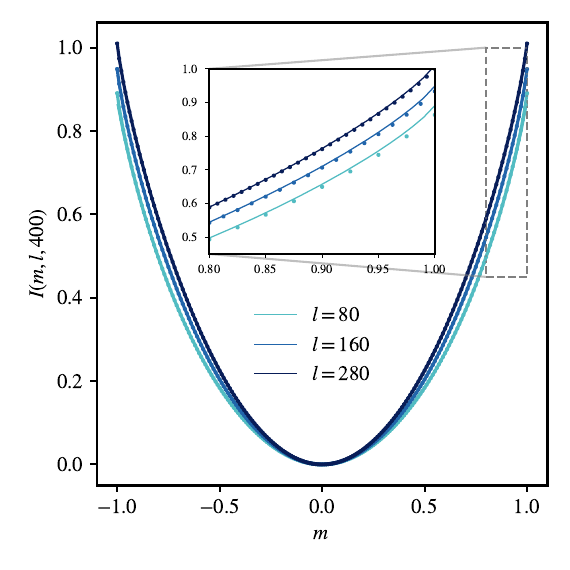}
    \caption{Comparison of rate function obtained through Legendre transformation of SCGF ( solid lines) and full probability distribution (dots) for a system with $N=400$ and multiple blocksizes $l$. The inset shows zoomed in region at the edge. $\Delta=0.4, \epsilon=1.0$ }
    \label{fig:compare_LT_prob}
\end{figure}

In this Appendix, we compare the rate function obtained from Legendre transformation of SCGF to direct estimation of the rate function from the full probability distribution.
\begin{equation}
I(m,l,N)=-\frac{1}{l} \ln \frac{P(m)}{P(0)}
\end{equation}
To obtain the full probability distribution for large blocksize $l$, we need to utilize high-precision calculation as the probability of extreme magnetization is exponentially suppressed. We compared the rate function estimated through both the Legendre transformation approach and full distribution for $\Delta=0.4, \epsilon=1.0$ and a system consisting of $N=400$ sites in Fig.~\ref{fig:compare_LT_prob}. For the full distribution data reported here, we carried out the calculation with a floating point precision of 600 bits (implemented using ArbNumerics library in Julia). The deviation between these two approaches is minimal. The deviation is maximum at the extreme $m$ values (shown in Inset of Fig.~\ref{fig:compare_LT_prob}).

\section{Ratio of dominant eigenvalues of transfer matrices\label{Appendix: dominant eigenvalues}}
\begin{figure}[H]
    \centering
    \includegraphics[width=\textwidth]{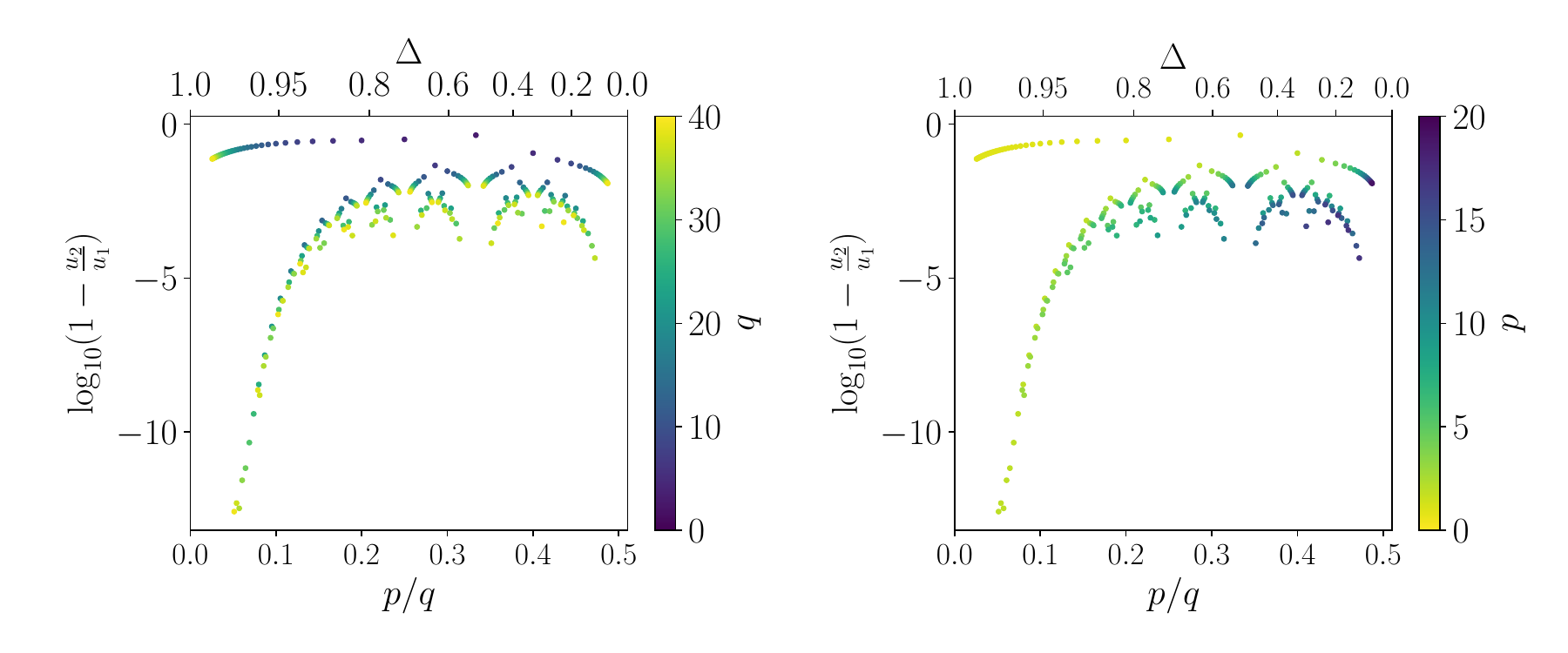}
    \caption{Log of deviation of ratio of  dominant eigenvalues of transfer matrix from $1$ corresponding to $U(\lambda)$ where $\lambda=0.6$ (See Sec.~\ref{SCGF}) and $\epsilon=1.0$. We look at a set of $\Delta$ which can be expressed as $\Delta=\cos(p\pi/q)$ with $q\leq 40$. The left panel has the points colored according to $q$ and the right panel has the points colored according to $p$. Similar results hold for general values of $\lambda$.}
    \label{fig:ratio_eigenvalues_V}
\end{figure}

In main text, we have shown the leading order correction to asymptotic-$N,l$ SCGF due to finite $l$ in eq.~\ref{leading_correction_scgf}. For $\Delta=\cos\frac{p}{q}\pi$, for rational values of $p/q$, convergence to large-$l$ SCGF is exponential in $l$ with the rate of convergence decided by the ratio of first two dominant eigenvalues of $U$ (See Sec.~\ref{SCGF}). Figure~\ref{fig:ratio_eigenvalues_V} shows the behavior of $\frac{u_2}{u_1}$ for a set of $\Delta$ values $\Delta=\cos(p\pi/q)$ where $p,q$ are co-prime integers less than $40$. 
The relative gap $1-u_2/u_1$ decreases with $p$ as shown in Fig.~\ref{fig:ratio_eigenvalues_V}(right) and for each value of $p$, the relative gap decreases with $q$  as shown in Fig.~\ref{fig:ratio_eigenvalues_V}(left).
The ratio goes down as $\Delta$ approaches $1$, thus a very slow convergence to asymptotic SCGF is expected in the Heisenberg limit suggesting large finite-size effects. 

\section{Finite-$N$ rate function \label{Finite-N-rate-function}}
\begin{figure}[htbp]
    \centering
    \begin{minipage}{0.48\textwidth}
        \centering
        \includegraphics[width=\linewidth]{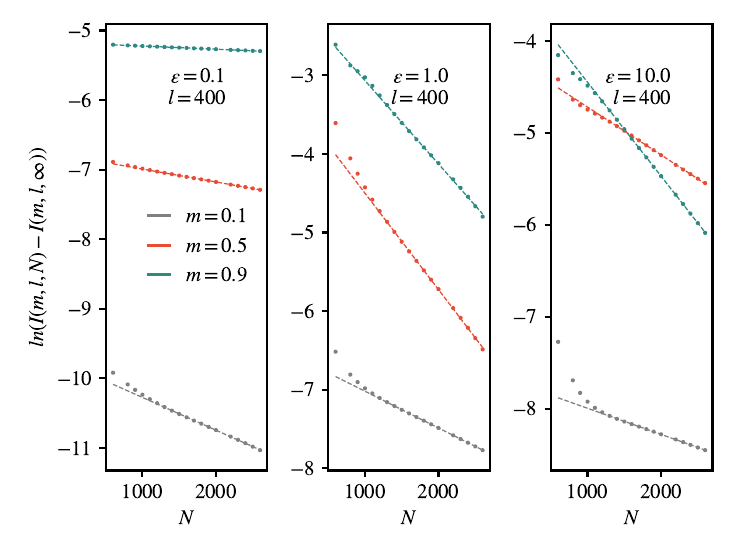}
        \caption{Scaling of rate function with $N$ at $l=400$ for $\Delta=0.2$ and multiple value of $\epsilon$ and $m$. The dots ar actual value obtained from simulation. The solid lines are fits obtained from Eq.~\ref{rate-function ansatz}.\label{fig:scaling_rate_function_N_asymptotic}}
    \end{minipage}
    \hfill 
    \begin{minipage}{0.48\textwidth}
        \centering
        \includegraphics[width=\linewidth]{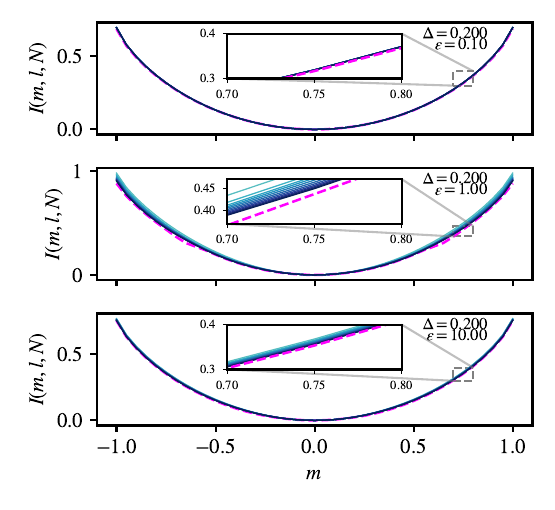}
        \caption{The rate function corresponding to the distribution of the average magnetization $m=M_z/l=1/l\sum \sigma_z$ in a block of size $l=400$ for NESS at $\Delta= 0.2 $ and $\epsilon=0.1,1.0$ and $10.0$. Different lines represent different $N$. The color grading goes from light green ($N=600$)) to blue ($l=2600$) in step of 200. The magenta dashed line is the asymptotic large-$N$ rate function obtained from Eq.~\ref{rate-function ansatz} at each $m$. (Inset) The zoomed in view of the main figure at extreme $m$.\label{fig:rate_function_N_asymptotic}}
    \end{minipage}
\end{figure}

To investigate the behavior of $I(m,l,N)$ with $N$, we take an ansatz for the rate function as follows:
\begin{equation}
I(m,l,N) \propto A e^{-bN}+I(m,l,\infty)
\label{rate-function ansatz}
\end{equation}
where $I(m,l,\infty)$ denotes the rate function at fixed $l$ and $m$ with $N \to \infty$. We test the validity of our scaling ansatz by investigating the convergence of $I(m,l,N)$ to large-$N$ asymptotic value, $I(m,l,\infty)$.In Fig.~\ref{fig:scaling_rate_function_N_asymptotic}, we verify our exponential scaling ansatz in Eq.~\ref{rate-function ansatz}. The data for $\Delta=0.2$ and multiple $l,\epsilon$ and $m$ is presented in Fig.~\ref{fig:rate_function_N_asymptotic}. The fitting parameters depend on $\delta,\epsilon$ and $m$. The fits clearly support the exponential ansatz for scaling of rate function with $N$ at fixed $l$.

For all cases, exponential convergence to large-$N$ asymptotic value is evident. The exponential convergence holds for all parameter regimes studied at large $N$ (only representative examples are presented here).

\end{document}